\documentclass{aa}  

\usepackage{graphicx}
\usepackage{natbib}
\usepackage[usenames]{color} 
\usepackage{amsmath}
\usepackage{amssymb}
\usepackage{listings}
\usepackage{adjustbox}

\usepackage{txfonts}
\newcommand{\matr}[1]{\mathbf{#1}}

\begin{document}

   \title{Spectral Line Identification and Modelling (SLIM) \\
   in the MAdrid Data CUBe Analysis (MADCUBA) package}
   \subtitle{An interactive software for data cube analysis}
   \titlerunning{SLIM and MADCUBA}
   
   \author{S. Mart\'in \inst{1,2}
          \and
          J. Mart\'in-Pintado\inst{3}
          \and
          C. Blanco-S\'anchez\inst{3}
          \and
          V. M. Rivilla\inst{4}
          \and 
          A. Rodr\'iguez-Franco\inst{5}
          \and 
          F. Rico-Villas \inst{3}
          }

   \institute{European Southern Observatory, Alonso de C\'ordova, 3107, Vitacura, Santiago 763-0355, Chile\\
              \email{smartin@eso.org}
         \and
             Joint ALMA Observatory, Alonso de C\'ordova, 3107, Vitacura, Santiago 763-0355, Chile
         \and 
Centro de Astrobiolog\'ia (CSIC-INTA), Ctra. de Torrej\'on a Ajalvir km 4, Madrid, Spain
         \and 
INAF-Osservatorio Astrofisico di Arcetri, Largo Enrico Fermi 5, 50125, Florence, Italy
         \and 
             Facultad de \'Optica y Optometr\'ia, Departamento de Biodiversidad, Ecolog\'ia y Evoluci\'on, Universidad Complutense de Madrid, Avenida de arcos de Jal\'on, 118, E-28037 Madrid, Spain
         }

   \date{}
 
  \abstract
   {The increase in bandwidth and sensitivity of state-of-the-art radio observatories is providing a wealth of molecular data from nearby star-forming regions up to high-z galaxies. Analysing 
   large data sets of spectral cubes requires efficient and user-friendly tools optimized for astronomers with a wide range of backgrounds.}
   {In this paper we present the detailed formalism at the core of the Spectral Line Identification and Modelling (SLIM) within the MAdrid Data CUBe Analysis (MADCUBA) package and their main data handling functionalities.
   These tools have been developed to visualize, analyze and model large spectroscopic data cubes.
}
   {We present the highly interactive on-the-fly visualization and modelling tools of MADCUBA and SLIM, which includes an stand-alone spectroscopic database. The parameters stored therein are used to solve the full radiative transfer equation under Local Thermodynamic Equilibrium (LTE). SLIM provides tools to generate synthetic LTE model spectra based on input physical parameters of column density, excitation temperature, velocity, line width and source size. SLIM also provides an automatic fitting algorithm to obtain the physical parameters (with their associated errors) better fitting the observations. Synthetic spectra can be overlayed in the data cubes/spectra to easy the task of multi-molecular line identification and modelling. 
}
   {We present the Java-based MADCUBA and its internal module SLIM packages which provide all the necessary tools for manipulation and analysis of spectroscopic data cubes. We describe in detail the spectroscopic fitting equations and make use of this tool to explore the breaking conditions and implicit errors of commonly used approximations in the literature. 
}
   {Easy-to-use tools like MADCUBA allow the users to derive the physical information from spectroscopic data without the need of resourcing to simple approximations.  SLIM allows to use the full radiative transfer equation, and to interactively explore the space of physical parameters and associated uncertainties from observational data.}


   \maketitle
%

\section{Introduction}
In the last decade, all large astronomical facilities operating at centimeter, millimeter and sub-millimeter wavelengths have continuously increased their sensitivity and instantaneous observing bandwidth, dramatically enhancing their potential scientific output. 
As examples of current bandwidth capabilities, the SMA now records instantaneously 32 GHz, NOEMA 15.4 GHz, IRAM~30m 15.6 GHz, and VLA 8 GHz. While ALMA currently records 8 GHz, it is planned to increase the bandwidth up to 16 GHz in the coming years\footnote{http://www.eso.org/sci/facilities/alma/announcements/20180712-alma-development-roadmap.pdf}. Future planed facilities also target wide instantaneous frequency coverage, such as the Next Generation VLA (ngVLA), expected to cover 20 GHz\footnote{http://ngvla.nrao.edu/page/refdesign} and the SKA 5 GHz\footnote{https://www.skatelescope.org/wp-content/uploads/2018/08/16231-factsheet-telescopes-v71.pdf}.  

The large bandwidths together with the superb sensitivities provided by both the new interferometric facilities and single-dish telescopes equipped with multi-beam receivers result in extremely large 3-dimensional spectroscopic data cubes. Such data sets contain huge amounts of physical, kinematic and chemical information in molecular and atomic recombination lines. 
This wealth of information allows us to study the physical conditions and the chemical processes taking place in a wide variety of astronomical objects, ranging from solar system objects and protoplanetary disks to galaxies observed in the early Universe.

While the instrumental capabilities have been improving, the tools to analyze the extremely rich and complex datasets are still in their infancy.
Using different strategies, during the last decade a number of tools have been developed in parallel for spectral line identification and molecular emission analysis at different levels. 
Here we briefly describe the tools available to date together with their key functionalities to serve as a comparison among them and with SLIM/MADCUBA as described below: 

$\bullet$ WEEDS\footnote{\url{https://www.iram.fr/IRAMFR/GILDAS/doc/html/weeds-html/weeds.html}} \citep{Maret2011}, developed in FORTRAN within the GILDAS\footnote{\url{https://www.iram.fr/IRAMFR/GILDAS/}} package, provides tools for the visualization of simulated molecular line profiles using the Local Thermodynamical Equilibrium (LTE) approximation superimposed on the observed single-pointing spectra. 

$\bullet$ CASSIS\footnote{\url{http://cassis.irap.omp.eu/}}, developed in Java, is a stand-alone package containing a molecular line catalog which is included in Herschel Interactive Processing Environment\footnote{\url{http://herschel.esac.esa.int/hipe/}} (HIPE) as a plug-in\footnote{Plug-in is software component that adds a specific tool to an existing package.}. It preforms both the LTE and the non-LTE analysis for single-pointing spectra. For the LTE analysis, CASSIS simulates the expected LTE molecular line profiles and visualize them on the observed spectra while relevant parameters can be  manually adjusted. It provides the LTE parameters and their associated errors derived by a non-linear least-squares fit to the data. For the non-LTE analysis, CASSIS connects to the Large Velocity Gradient (LVG) code RADEX\footnote{\url{http://home.strw.leidenuniv.nl/~moldata/radex.html}} (\citealt{vandertak2007}) when collisional cross sections for the molecular species are available, providing the H$_2$ densities and the molecular column densities for an assumed kinetic temperature.

$\bullet$ XCLASS\footnote{\url{https://xclass.astro.uni-koeln.de/}} \citep{Moeller2017}, mainly developed in FORTRAN, can be executed within CASA\footnote{\url{https://casa.nrao.edu/}} \citep{McMullin2007} and performs the LTE line profile simulation and the non-linear least-squares fit to the data from both single pointing spectra and 3D spectroscopic data cubes. It allows to have a very complex distribution of molecular and continuum slabs (clouds) along the line of sight, considering the effects of the molecular line attenuation by foreground dust and the molecular line absorption of the background continuum. This tool runs with python scripts using input files and delivers the results of fitted LTE parameters and their associated errors. 

With the advent of ALMA and its objective of providing science-ready data products, molecular astrophysics and astrochemistry are now becoming unique and very powerful tools within the grasp of non-experts in radio/(sub)millimeter astronomy and molecular astrophysics. 
In the past, the analysis of the molecular data was based on a number of approximations helping to simplify the derivation of physical conditions and molecular abundances from very limited set of data. Usually these approximations have a limited range of applicability and in many cases provide degenerated results. Thanks to current facilities such as ALMA, it is now possible to choose the appropriate set of molecular transitions that when combined with suited tools will not require to rely on such strong approximations. This combination allows the user to better constrain the physical parameters by removing some of the degeneracies. 

Newly developed analysis tools must also meet the challenging requirement of being intuitive and easy-to-use, as well as easily and efficiently handling large volumes of data. 
The MAdrid Data CUBe Analysis package (MADCUBA), developed at the Centro de Astrobiolog\'ia (CSIC, INTA) has been designed relying on two basic premises: efficiency (required to handle large and complex data sets) and easiness of use, thanks to user-friendly intuitive interfaces for non-experts on molecular astrophysics. 

In this paper we present and briefly describe the main data handling functionalities of MADCUBA. In summary, and for the sake of comparison with the existing tools described above, MADCUBA allows interactive manipulation of single-pointing spectra and  spectroscopic data cubes with a number of functions working in both spatial and spectroscopic axes. The Spectral Line Identification and Modeling (SLIM), at the core of MADCUBA, provides spectral analysis tools for modelling molecular emission. 
Here, we provide details on the formalism used SLIM, where the equations, approximations and limitations are {\bf comprehensively} described.
SLIM provides molecular and recombination line identification tools as well as LTE molecular analysis for emission and absorption lines for both single-pointing spectra and spectroscopic data cubes. A background continuum source can be added to produce absorption profiles, but attenuation by dust is not yet implemented. Such analysis tools are not available for recombination lines within SLIM yet. However, both the dust attenuation and the recombination line modelling will be implemented in the future. The generation of synthetic LTE molecular spectra is fully interactive through sliders allowing users to change the input physical parameters, with on-the-fly visualization of the resulting model spectra overlaid on the observed data/data cubes.
Besides, it provides tools to automatically fit physical parameters with their associated errors derived from non-linear least-squares fit to the data. 

MADCUBA has already been used in the last years in many papers addressing different scientific topics in a variety of astronomical sources: prestellar cores (\citealt{Jimenez-Serra2016}), infrared-dark clouds (\citealt{Cosentino2018}), low-mass star-forming regions (\citealt{Martin-Domenech2017,Rivilla2019}), high-mass star-forming regions (\citealt{Rivilla2016,Rivilla2017,Rivilla2017a,Rizzo2017,Zahorecz2017,Colzi2018,Moscadelli2018,Beltran2018}), Galactic Centre giant molecular clouds (\citealt{Zeng2018,Rivilla2018,Riquelme2018,Rivilla2019a}) and HII regions (\citealt{Armijos-Abendano2018}), and external galaxies (\citealt{Martin2014,aladro2015,Martin2015,Harada2018,Sewilo2018,Martin2019}). 

The paper is organized as follows:
In Sec.~\ref{sec.madcuba} we enumerate the main cube handling functionalities of MADCUBA. 
In Sec.~\ref{sec.slim} we describe in detail the SLIM tool. We present the spectroscopic database, the basic radiative transfer formulation and the assumptions made to generate synthectic molecular spectra in LTE. We also explain the automatic non-linear least-squares fitting algorithm and discuss the derived errors associated to the fitted LTE parameters and the quality of the fit.
We present a number of examples to illustrate the potential of SLIM. 
To go beyond a mere presentation of the radiative transfer LTE formulation, in Sect.~\ref{sec.approximations}, we discuss common approximations used in the literature to analyze molecular spectra and point out the limitations and degeneracies of such approximations. Building on this discussion, we describe how MADCUBA can allow the estimate of the uncertainties derived from these degeneracies. 
  

\section{MAdrid Data CUBe Analysis package (MADCUBA)}
\label{sec.madcuba}

MADCUBA\footnote{\label{madcubaweb}\url{http://cab.inta-csic.es/madcuba/MADCUBA_IMAGEJ/ImageJMadcuba.html}} is a package developed to import, visualize, manipulate, process and analyze molecular and recombination line astronomical data from both 3D spectroscopic cubes and single-pointing spectra.
It has been designed to combine an user-friendly interface and a powerful visualization and data analysis system able to deal with large volumes of data. MADCUBA provides simple tools to analyze and interpret molecular and recombination line spectroscopic observations.
It is is a stand-alone package, developed in Java\footnote{\url{https://www.java.com}} as a plug-in for ImageJ\footnote{\url{https://imagej.net/}; \url{https://imagej.net/ImageJ}} \citep{Schneider2012,Schindelin2015}.
ImageJ is a Java-based highly extensible open source image processing program which has been developed by the U.S. National Institutes for Health for the analysis scientific multidimensional images mainly in Biology. ImageJ provides the core infrastruture for the visualization and processing of data cubes, and the framework for interactive scripting. 
Additionally, MADCUBA makes use of the STIL\footnote{Starlink Tables Infrastructure Library \url{http://www.star.bris.ac.uk/~mbt/stil/}}, non-tam-fits \footnote{\url{http://nom-tam-fits.github.io/nom-tam-fits/}}, and skyview Libraries\footnote{\url{https://skyview.gsfc.nasa.gov/current/cgi/titlepage.pl}}.
STIL is a library used for managing large tables, non-tam-fits is a Java library for reading and writing FITS (Flexible Image Transport System) files, and Skyview provides the spatial World Coordinate System infrastruture.

MADCUBA is an out-of-the box\footnote{Ready-made software that works without any special configuration, installation or modification.} toolkit with no other requirements than Java 1.8 or above. It can be downloaded from the MADCUBA website$^{\ref{madcubaweb}}$  and does not require installation. After decompressing the downloaded file, it can be run directly from the executable files within the MADCUBA directory.
It is compatible with Linux, Mac OS X, and Windows environments.
The package can be easily updated to the latest version by replacing two .jar files that can be independently downloaded from the website.

\subsection{Importing data into MADCUBA}

MADCUBA can read/import data from the main radio to far-IR observatories in the world (i.e. ALMA, Herschel, NOEMA, SMA, VLA, GBT, IRAM~30m, Effelsberg~100m). The native data format for both data cubes and spectra is FITS.  MADCUBA does import single spectra and data cubes in FITS format generated by CASA, GBT-IDL, MIRIAD. Herschel products (versions 2, 2.5 and 3) from the Herschel Science Archive\footnote{\url{http://archives.esac.esa.int/hsa/whsa/}} for the 3  instruments (PACS, SPIRE and HIFI) and GILDAS single spectra (a working local GILDAS installation is required) and GILDAS data cubes (previously exported to FITS) can also be imported. In addition, MADCUBA offers a generic data cube import option, able to deal with a large variety of data cubes parameters, as well as the possibility of importing spectra from a plain formatted ASCII file (see Appendix ~\ref{sec.sampleasciiformat} for a sample input file).



\subsection{Visualization and handling of data cubes and spectra}
\label{sec.cubevisualization}

\begin{figure*}
\includegraphics[width=\textwidth]{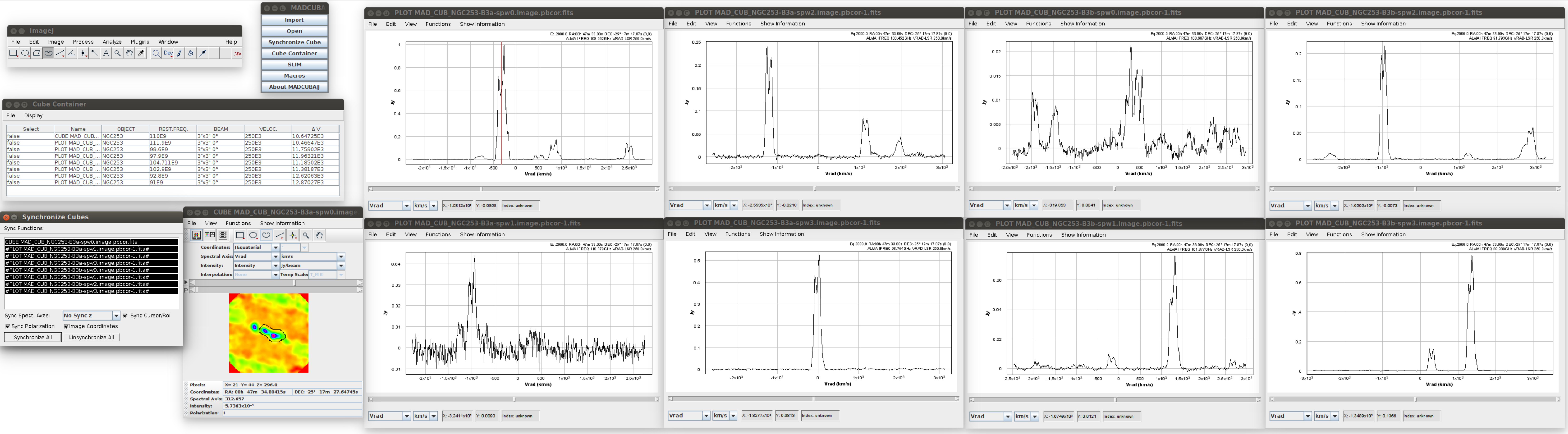}
\caption{Screenshot showing the use of MADCUBA to spatially synchronize multiple datacubes. In this case eight ALMA datacubes taken from the data published by  \citet{Martin2019} are loaded and synchronized. From top to bottom and left to right, the windows shown are: the main ImageJ window; the main MADCUBA menu; the cube container window displaying the currently opened data cubes; the synchronize cube window listing the data cubes currently synchronized (in this case all opened ones); the cube display where the selected channel of the first opened data cube is displayed and the RoI is interactively defined (black contour on image); and finally the eight spectra corresponding to the integrated spectra from the RoI for each synchronized datacube. Note that the image channel displayed is selected with a red line in the top left spectrum, which corresponds to the spectrum from the cube displayed.
}
\label{fig.MADCUBAscreenshot}
\end{figure*}

MADCUBA uses the powerful infrastructure offered by ImageJ to interactively visualize and handle astronomical data cubes and spectra through a graphical user interface (GUI).
MADCUBA allows interactive visualization of single/multiple data cubes and spectra. In Fig.~\ref{fig.MADCUBAscreenshot} it is shown an screenshot with a sample MADCUBA session where its main windows are displayed and identified.

The spectral visualization offers the possibility to interactively convert both the spectral and intensity axes to the most commonly used units in radio, mm, sub-mm, and far-IR astronomy (see Appendix~\ref{sec.units}). The visualization allows to merge and/or overlay multiple spectra in a single plot when the spectral axis is set to frequency, wavelengths, or energy.  

The cube visualization is achieved by on-the-fly displaying the image plane at the selected spectral channel in the cube, and plotting the spectrum from the selected spatial pixel(s) in the image plane. The latter includes the possibility to show the integrated spectrum over a spatial region of interest (RoI) defined on the image as a rectangle, ellipse, polygon, or line. 
Similarly to other common visualization packages such as ds9\footnote{\url{http://ds9.si.edu/site/Home.html}} \citep{Joye2003}, RoIs can be defined on the image with different geometries (rectangle, ellipse, polygon, or line). RoIs definition can be performed either interactively or through command line scripting (Sect.~\ref{sec.scripting}). The defined RoI is recorded in the cube history file and, like ds9, it can be loaded onto the image from the ImageJ scripting window. The cube visualization also allows to change spectral and intensity units and to choose between equatorial and Galactic coordinates. The spectrum in the plot can be extracted and saved for further analysis. The intensity/flux units and the solid angle corresponding to the selected RoI of the extracted spectra are calculated and saved to the spectra accordingly. These parameters will be used for modelling within SLIM (Sect.~\ref{sec.slim}). 

In addition to single cube visualization, MADCUBA allows spatially-synchronized visualization of multiple cubes. The pixel or RoI selected in a cube is propagated to all the synchronized cubes in celestial coordinates. The propagated RoIs are displayed on the images of all synchronized cubes and the corresponding spatially integrated spectra are plotted (see Fig.~\ref{fig.MADCUBAscreenshot}). The synchronized multiple spectra from all the cubes can also be extracted and saved for further analysis with SLIM (Sect.~\ref{sec.slim}). 
This is particularly handy for spectral line surveys carried out via multiple frequency tunings or even a simple ALMA project whose products will contain multiple cubes of individual spectral windows.

MADCUBA also offers a number of tools for data cube and spectra manipulation such as spectral baseline for cubes and single spectrum, spatial and spectroscopic smoothing and cropping, spectral interpolation and re-sampling, Gaussian fit to line profiles, and velocity/frequency integrated images from cubes. Cropping and smoothing can be also applied simultaneously to synchronized data cubes.


\subsection{Scripting and history file}
\label{sec.scripting}
ImageJ contains a plethora of contributed scripts and plug-ins, and allows for user-scripting in Jython\footnote{\url{https://www.jython.org/index.html}}.
MADCUBA makes use of the infrastructure of the  Macro/scripting  language (IJM\footnote{see \url{https://imagej.nih.gov/ij/developer/macro/macros.html} for a description of the potential of IJM}) built into ImageJ that allows controlling many aspects of MADCUBA and ImageJ. Most of the functions and tools in MADCUBA can be used in scripts as a sequence of actions to manipulate large data sets of cubes and spectra in an automatic way (see Appendix~\ref{sec.samplemacro} for a sample script). Scripts can be easily written in MADCUBA since the operations performed by the user can be recorded automatically. The recorded scripts can be then edited and tested interactively to apply the same operations to other data sets. Frequently used scripts can be installed in MADCUBA to be used as tools, and they can also be distributed to other users. 

The MADCUBA product is composed of two files, the data file and the history file. The history file contains all the operations performed to the data and is written in the scripting language. This file allows for easily tracking and repeating all the operations performed to the data and/or to edit the parameters used in the data processing.


\section{Spectral Line Identification and Modeling (SLIM)}
\label{sec.slim}
\begin{figure*}
\includegraphics[width=\textwidth]{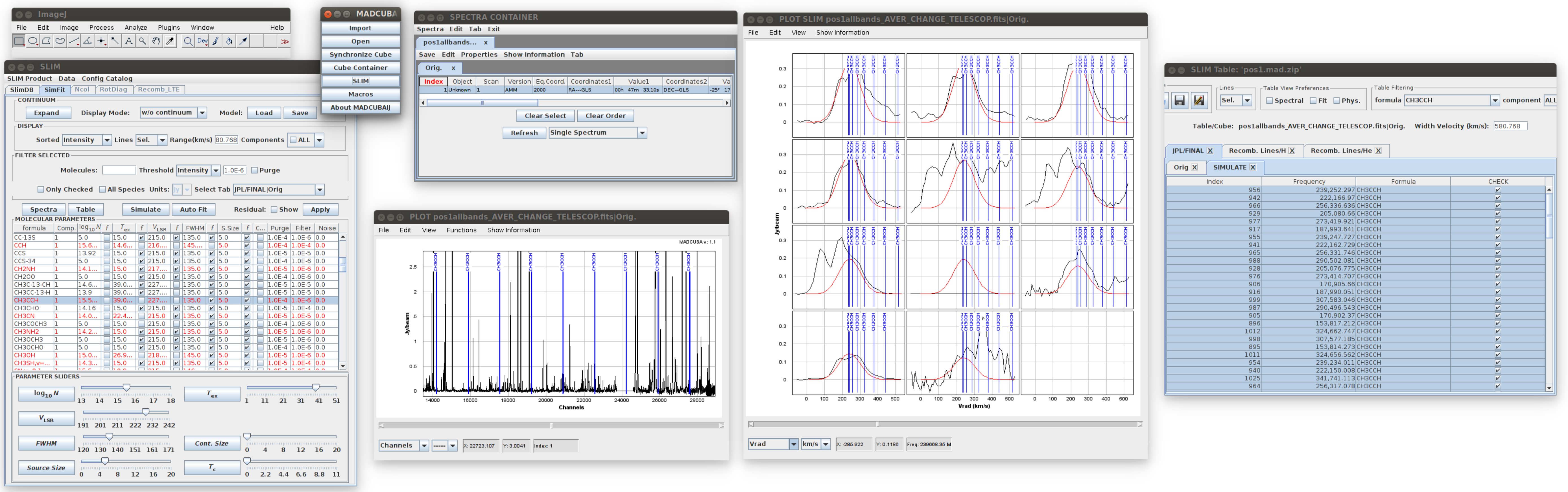}
\caption{Screenshot showing the use of SLIM within MADCUBA to identify and fit a synthetic LTE model to an observed broad band spectrum. In this case a 30000 channel ALMA spectrum is displayed where more than 100 molecular species have been fitted. From top to bottom and left to right, the windows shown are: the main ImageJ window; the main MADCUBA menu; the modelling tab within the SLIM tool with the list of fitted species and the sliders in the bottom to interactively adjust the physical parameters; the spectra container window displaying the currently opened spectra; the plot showing the full spectrum with the transitions of the selected species labeled; the SLIM PLOT showing the synthetic model overlaid on top of the observed spectra zoomed to the brightest transitions of the selected species; the list of transitions of the selected species.
}
\label{fig.SLIMscreenshot}
\end{figure*}

SLIM is a key module of MADCUBA for the analysis of spectra (Fig.~\ref{fig.SLIMscreenshot}).
SLIM tools allow for line identification of spectroscopic features through queries to its spectroscopic database (Section~\ref{sec.database}) which can be overlaid on top of the displayed spectra. It also provides the infrastructure to model the molecular emission (Sect.~\ref{sec.LTE}) and to generate synthetic spectra from input physical parameters (Section~\ref{sec.syntheticspectra}). These synthetic spectra can be overlaid on top of the observed spectra and/or fitted to the observations in order to derive the best fit to the physical parameters with their corresponding uncertainties (Section~\ref{sec.fitting}).
An extension of this tool to automatically analyze data cubes is under development. 

\subsection{Spectroscopic Database}
\label{sec.database}

Spectroscopic information is available off-line within SLIM through a stand-alone HSQLDB (HyperSQL DataBase\footnote{http://hsqldb.org/}). 
The database contains the following default spectral line catalogs:
\begin{itemize}
\item[\textbullet]{Jet Propulsion Laboratory (JPL\footnote{\url{http://spec.jpl.nasa.gov/ftp/pub/catalog/catform.html}}) \citep{Pickett1998}}
\item[\textbullet]{Cologne Database for Molecular Spectroscopy (CDMS\footnote{\url{http://www.astro.uni-koeln.de/cgi-bin/cdmssearch}}) \citep{Muller2001,Muller2005,Endres2016}}
\item[\textbullet]{Spectral Line Atlas of Interstellar Molecules (SLAIM) \citep{Lovas1984}}
\item[\textbullet]{Recommended Rest Frequencies for Observed Interstellar Molecular Microwave Transitions, 1991 revision \citep{Lovas1992}}
\item[\textbullet]{Recommended Rest Frequencies for Observed Interstellar Molecular Microwave Transitions, 2002 revision \citep{Lovas2004}}
\item[\textbullet]{Radio recombination line spectroscopy database calculated from \citet{Towle1996} by \citet{Baez2014}}
\end{itemize}

The first two entries (JPL and CDMS) contain the needed spectroscopic parameters to be used as input for simulating and fitting {\bf molecular} spectra using LTE analysis (Sect.~\ref{sec.syntheticspectra}). More precisely, for each transition, the spectroscopic databases provide the rest frequency in MHz, the integrated  intensity  in  units  of $\rm nm^2 MHz$ at 300~K, and the lower state energy in $cm^{-1}$. Together with the partition function of the molecule for a defined set of temperatures, also provided in the databases, we derive the Einstein coefficient $A_{ul}$ which are used for estimating the line optical depth (Sect.~\ref{sec.lineprofiles}).

The last four entries only include the species, transitions and rest frequencies used exclusively for line identification. While SLAIM theoretical catalog is included for the sake of completeness, the catalog of observed interestellar molecules can be actually handy for quick identification of transitions already identified in the ISM. Modelling tools for recombination lines are not available yet, but will be implemented in future releases.

The SLIM database is regularly updated based on the new entries of the CDMS and JPL catalogs. SLIM also allow the user to upload new molecules not included in these catalogs if the spectroscopy and the partition function are provided in CDMS/JPL format.





\subsection{Molecular emission in Local Thermodynamical Equilibrium (LTE)}
\label{sec.LTE}

Derivation from first principles of the expression for the molecular column density as a function of spectral line observables can be found in e.g. \citet{Mangum2015}.
In this section, we describe the basic radiative transfer formalism as well as the assumptions made by MADCUBA to calculate the synthetic spectra of molecular species used to derive the physical and kinematic properties of the emitting regions. The main assumptions and the equations where these assumptions are implicit are the following:
\begin{itemize}
\item[\textbullet]{Uniform temperature and density slab of emitting gas (Eq.~\ref{eq.radiativeTransfer})}
\item[\textbullet]{Emitting regions are circular Gaussian shaped (Eq.~\ref{eq.unitssizes})}
\item[\textbullet]{Optical depth Gaussian profiles (Eq.~\ref{eq.taushape})}
\item[\textbullet]{Local Thermodynamic Equilibrium (Eq.~\ref{eq.LTE})}
\end{itemize}

\subsubsection{Radiative transfer formalism}
\label{sec.radiativetransfer}
 The  intensity of a source ($I_\nu$) can be expressed through the Planck function, $B_\nu(T_B)$, as a function of the brightness temperature, $T_B$:
\begin{equation}
I_\nu=B_\nu(T_B) ~=
~\frac{2k\nu^2}{c^2} J_\nu(T_B)
\label{eq.planck}
\end{equation}
where $J_\nu$ is the intensity in temperature units:
\begin{equation}
J_\nu(T)=\frac{h\nu/k}{e^{h\nu/kT}-1}
\label{eq.intensityinT}
\end{equation}

We define the Radiation temperature as $T_R=J_\nu(T_B)$, which through Eq.~\ref{eq.planck} is proportional to the intensity as





\begin{equation}
I_\nu= \frac{2k\nu^2}{c^2} T_R 
\label{eq.radiationtemperature}
\end{equation}
The radiation temperature will be equal to the brightness temperature in the  Rayleigh-Jeans (R-J) approximation when $h\nu/kT_{ex}\ll 1$. Since this condition does not hold for low temperatures and high frequencies, SLIM does not use the R-J approximation.

SLIM considers the solution of the radiative transfer equation assuming an uniform (temperature and density) slab of gas. Along the line of sight, we consider that in addition to the Cosmic Microwave Background with a brightness temperature $T_{bg}$ there may be a background continuum source with a temperature $T_c$. For this case, the solution of the standard radiative transfer equation for a slab of a given molecular species with an excitation temperature ($T_{ex}$) can be written, as a function of the frequency, as:

\begin{equation}
T_{R} = J_\nu(T_{bg}) ~ e^{-\tau_\nu} + 
~ c_s~J_\nu(T_{c}) ~ e^{-\tau_\nu} + J_\nu(T_{ex}) ~ (1-e^{-\tau_\nu})
\label{eq.radiativeTransfer}
\end{equation}

where the total frequency dependent optical depth $\tau_\nu$ (Sect.~\ref{sec.lineprofiles}) is the sum over the opacity of all the transitions as:

\begin{equation}
\tau_\nu = \sum_t \tau_\nu^t,
\label{eq.sumtau}
\end{equation}



and the covering factor $c_s$ considers the fraction of continuum absorbed by the foreground gas. This is the fraction of the solid angle of the continuum source ($\Omega_c$)  which is covered by the molecular cloud, and therefore cannot be larger than unity. 
As explained in Sect.~\ref{sec.syntheticmultimolecule}, outside the frequency range of the emitting lines, the emission will be that of the whole continuum source and background.

Since we are interested only in the intensity of the line,  the continuum 
subtracted line intensity, $T_L$ is obtained by subtracting the continuum contribution $J_\nu(T_{bg}) + c_s~J_\nu(T_{c})$ to Eq.~\ref{eq.radiativeTransfer} as:
\begin{equation}
T_L=\bigl( J_\nu(T_{ex})-J_\nu(T_{c})-J_\nu(T_{bg}) \bigr) (1-e^{-\tau_\nu})
\label{eq.TlineradiativeTransfer}
\end{equation}

Note that for the sake of simplicity, in this equation it is implicit a line covering factor $c_s=1$ since, as it will be described in Sect.~\ref{sec.syntheticlinearsuperposition}, for modeling purposes the line emission can be divided in components such that the absorbing component satisfies this assumption. The validity of this assumption will be discussed in Sec.~\ref{sec.continuum} when considering continuum observable units. 






\subsubsection{Observed quantities. Source sizes}
\label{sec.observables}

The intensity of the line, $T_L$, in Sect.~\ref{sec.radiativetransfer} refers to the radiation emitted at the surface of the slab. However we are interested in the detected line intensity when observed with a radio telescope. The actual observed intensity will be then the result of the convolution of the telescope spatial response with the radiation temperature distribution of the emitting source. Thus the line (or radiation) temperature can be related to the observable quantities of flux density $S$ or main beam brightness temperature $T_{\rm MB}$ as
\begin{equation}
S = \frac{2k\nu^2}{c^2} \int T_L ~\delta\Omega_{s}  =  \frac{2k\nu^2}{c^2} \int T_{MB} ~\delta\Omega_{s\star b}
\label{eq.fluxdensity}
\end{equation}

Where $\Omega_s$ is the solid angle of the source, and $\Omega_{s\star b}$ is the solid angle of the convolution of source and the main beam or point spread function of the telescope, which will be the solid angle of the main beam $\Omega_{b}$ in the case of point sources \citep[see][for details]{Downes1989}. 


Based on Eq.~\ref{eq.fluxdensity}, SLIM does calculate the spectrum in either temperature ($\rm K$) or flux density units ($\rm Jy$) from the line temperature in Eq.~\ref{eq.TlineradiativeTransfer} as
\begin{equation}
\begin{array}{l}
T_{\rm MB}= \frac{\Omega_s}{\Omega_{s\star b}} ~T_{L} 
\\
\\
S=10^{23} ~\frac{2k\nu^2}{c^2} ~\Omega_{s} ~T_{L} \\
\end{array} 
\label{eq.units}
\end{equation}

where $10^{23}$ is the conversion factor to Jy in cgs units, and $\Omega$ are in $steradians$. Data in other supported intensity units in MADCUBA are internally converted to any of these two basic units (see Appendix~\ref{sec.units}). 

Assuming that both the source and the telescope beam are represented by 2D elliptical Gaussian functions, the solid angle is calculated as $\Omega=\frac{\pi}{4ln(2)}\theta_{major}\theta_{minor}=1.133~\theta_{major}\theta_{minor}$ where $\theta_{major}$ and $\theta_{minor}$ are the full width at half maximum (FWHM) of the ellipse major and minor axis, respectively. Since the source morphology within the beam of the telescope is unknown, SLIM considers, for simplicity, that both the source and the beam to be circular. The beam  size in SLIM is defined as the FWHM of a circular Gaussian  with the same solid angle than an elliptical beam, $\theta_{b} = \sqrt{\theta_{major} \theta_{minor}}$. The convolution of the beam and the source is then another Gaussian with a FWHM=$\sqrt{(\theta_{s}^2 + \theta_{b}^2)}$ ~ and the convolved solid angle $\Omega_{s\star b}=1.133~(\theta_{s}^2 + \theta_{b}^2)$. 

Under these assumptions, Eq.~\ref{eq.units} can be rewritten in terms of source, and beam FWHM as

\begin{equation} 
\begin{array}{l}
T_{\rm MB}= \frac{\theta_{s}^2 }{\theta_{s}^2 +\theta_{b}^2} ~ ( J_\nu(T_{ex}) -  J_\nu(T_{c}) - J_\nu(T_{bg})) ~ (1-e^{-\tau_\nu})  \\
\\
S=10^{23} ~\frac{2k\nu^2}{c^2} 1.133~\theta_{s}^2 ~ ( J_\nu(T_{ex}) -  J_\nu(T_{c}) - J_\nu(T_{bg})) ~ (1-e^{-\tau_\nu})  \\
\end{array} 
\label{eq.unitssizes}
\end{equation}

In the case of extended sources, the total flux density of the source will be that within a beam (i.e. $\theta_{s} = \theta_{b}$ for $\theta_{s} > \theta_{b}$).



\subsubsection{Background continuum. Absorption lines }
\label{sec.continuum}
In the presence of a background continuum source the slab of gas will not only emit, but it could also absorb continuum photons  producing both absorption and emission line profiles. Eq.~\ref{eq.TlineradiativeTransfer} shows that absorption will occur when  $J_\nu(T_{ex}) < J_\nu(T_{c})+J_\nu(T_{bg})$. SLIM provide several ways to introduce the continuum effects in the radiative equation. 
The background continuum can be provided by the user through an input ASCII file, with comma separated pairs of rest frequencies (in Hz) and continuum intensities in the corresponding units, covering the whole frequencies range of the transitions to be simulated (see Appendix~\ref{sec.samplecontinuuminput} for a sample input file). The continuum can also be extracted from the data when observations contain both line and continuum emission. This is the case for interferometric observations or single dish special observing modes. The baseline functionality (mentioned in section \ref{sec.cubevisualization}) can be used to fit and remove the continuum from the spectra. The subtracted baseline will be considered by SLIM as the continuum background when required. 
In addition to these options to define the continuum level based on input from observations, SLIM also provides the following models for the continuum emission that will fit most of the observed continuum spectra arising from synchrotron, free-free and dust emission. The continuum model options are:

\begin{itemize}


\item[\textbullet]Cosmic Microwave background (CMB) at temperature  $T_{bg}$ that defaults to 2.73 K: 
\begin{equation}
{ 
J_\nu(T_{bg}) = \frac{h\nu}{k} \frac{1}{e^{h\nu/kT_{bg}}-1}, 
}
\label{eq.bg}
\end{equation}

\item[\textbullet]Black body at a temperature  $T_{bb}$ and size $\theta_{c}$

\begin{equation}
{ 
J_\nu(T_{bb}) = 
\frac{h\nu}{k} \frac{1}{e^{h\nu/kT_{bb}}-1} , 
}
\label{eq.bb}
\end{equation}

\item[\textbullet]
Modified black body, at a temperature $T_{gb}$ and size $\theta_{c}$ to model dust emission
\begin{equation}
{ 
J_\nu(T_{gb}) = 
J_\nu(T_{bb})(1 - e^{-\tau(\nu)}) , 
}
\label{eq.gb}
\end{equation}

where the blackbody at $T_{bb}$, 
is modified 
for the lower frequencies 
with an optical depth 
$\tau(\nu)$=$\tau_{\rm 0}(\nu/\nu_{0})^{\beta}$.
The dependence of the optical depth with frequency is related to the dust absorption coefficient, where $\beta$ is the dust emissivity index.
%
%

\item[\textbullet]Power law with size $\theta_{pw}$ and size $\theta_{c}$ to model synchrotron, ionized winds and free-free emission

\begin{equation}
{ 
J_\nu(T_{pw}) = 
J_{\rm pw} (\nu/\nu_{\rm pw})^{\alpha_{\rm pw}}
}
\label{eq.pw}
\end{equation}
which requires an input intensity ($J_{\rm pw}$) at a frequency $\nu_{\rm pw}$, and a spectral index $\alpha_{\rm pw}$.



\end{itemize}

Similar to the relations in Eq.~\ref{eq.unitssizes}, continuum emission will be then calculated in the corresponding units as
\begin{equation} 
\begin{array}{l}
T_{\rm MB}^{cont}= \frac{\theta_{c}^2 }{\theta_{c}^2 +\theta_{b}^2} ~ J_\nu(T_{c}) \\
\\
S^{cont}=10^{23} ~\frac{2k\nu^2}{c^2} 1.133~\theta_{c}^2 ~ J_\nu(T_{c}) \\
\end{array} 
\label{eq.unitssizescontinuum}
\end{equation}

for the units of main beam temperature and flux density, respectively.

We note that, under the assumption of continuum with a circular Gaussian distribution with a size $\theta_{c}$, then the covering factor in Eq.~\ref{eq.radiativeTransfer} can be written as 
$c_s=\theta_{s}^2/\theta_{c}^2$.
We can then write the continuum contributing to the line emission as
\begin{equation} 
\begin{array}{l}
T_{\rm MB}^{cont}= \frac{\theta_{c}^2 }{\theta_{c}^2 +\theta_{b}^2} ~ \frac{\theta_{s}^2 }{\theta_{c}^2}~J_\nu(T_{c})
\sim
\frac{\theta_{s}^2 }{\theta_{s}^2 +\theta_{b}^2} ~J_\nu(T_{c}) ~, ~ {\rm for}~ \theta_{s}\sim\theta_{c} \\
\\
S^{cont}=
10^{23} ~\frac{2k\nu^2}{c^2} 1.133~\theta_{c}^2 ~ \frac{\theta_{s}^2 }{\theta_{c}^2}~ J_\nu(T_{c}) 
= 
10^{23} ~\frac{2k\nu^2}{c^2} 1.133~\theta_{s}^2  J_\nu(T_{c}) 
\\
\end{array} 
\label{eq.unitssizescontinuumcontributiontoline}
\end{equation}
Thus, the continuum contributing to the radiative transfer in the line is equal, or approximately equal in the case of temperature units, to the continuum emission coming from a region equal to the size of the source emitting region $\theta_s$, which is used in Eq.~\ref{eq.TlineradiativeTransfer}.

For the cases of in which the user selects a continuum from fitted baseline or continuum spectrum from input file, it will be necessary to explicitly calculate the continuum fraction that is contributing to the line emission. For this purpose,
the covering factor will be defined as $c_s=\theta_{s}^2/\theta_{c}^2$, where $\theta_c$ is the size of the emitting continuum source, that can be input in SLIM, and $\theta_{s}$ is that of the molecular cloud.
In all cases, as explained in Sec.~\ref{sec.syntheticmultimolecule}, the continuum covering factor is limited to 1 (Sec.~\ref{sec.radiativetransfer}), and thus $\theta_{s} \leq \theta_{c}$).

All the continuum models can be combined to fit the observed continuum spectra. SLIM can display the continuum emission underneath the line spectrum allowing to visualize the fit to the continuum spectrum on-the-fly by changing with slicers the continuum parameters of temperature, $\beta$ and size. The continuum model is defined for all molecules observed in a give position of the sky. 

The effect produced by the continuum emission as background can be selected individually per molecule and/or velocity component to be considered in the radiative transfer equation (see Sect.~\ref{sec.syntheticmultimolecule}).

\subsubsection{Line profile and optical depth}
\label{sec.lineprofiles}

SLIM assumes that the small scale structure motions of gas particles in the slab is characterized by a Maxwellian velocity distribution independent from its physical origin being thermal and/or turbulent. In this case, the velocity/frequency dependence of the line opacity profile will be described by the Gaussian function
\begin{equation}
\tau_{\rm v}=\tau_\circ e^{-4\,ln 2(\rm v-v_\circ)^2/\Delta v^2}
\label{eq.taushape}
\end{equation}
where $\tau_\circ$ is the opacity at the central velocity ($\rm v_o$) of the line profile. $\Delta \rm v$ is the full width at half maximum that can be related to the variance $\sigma^2$ of the Gaussian as  $\Delta \rm v = 2\sqrt{2~ln2} \sigma = 2.354 \sigma$. Non-Gaussian observed profiles will have to be modelled as a linear combination of Gaussian velocity components (Sect.~\ref{sec.syntheticlinearsuperposition})

When the optical depth frequency dependence from Eq.~\ref{eq.taushape} is included  in Eq.~\ref{eq.unitssizes} it results in Gaussian profiles for low opacities (optically thin regime), but saturated flat top profiles for high opacities (optically thick regime) as shown in the sample spectral profiles in Fig.~\ref{fig.RotationalDiagram}. Just to note that, from Eq.~\ref{eq.TlineradiativeTransfer}, line profiles will saturate at  $T_{L}\sim T_{ex}-T_{c}-T_{bg}$ ~ in the Rayleigh-Jeans approximation and assuming an extended source size covering the continuum source.


The integral of the opacity function over the velocity distribution of the gas particles for a given transition can be related to physical parameters of column density of gas particles in the upper level of the transition $N_u$, the excitation temperature $T_{ex}$, and the velocity dispersion $\Delta v$  as
\begin{equation}
\int{\tau_{\rm v} \delta \rm v}~=~ 1.064467~\tau_\circ ~ \Delta {\rm v}~=~\frac{c^3}{8\pi\nu_{ul}^3} A_{ul} N_u (e^{h\nu_{ul}/kT_{ex}}-1)
\label{eq.tau}
\end{equation}
where 1.064467 is just a geometrical factor based on the assumed Gaussian profile as described below, $A_{ul}$ is the Einstein coefficient, and $\nu_{ul}$ the frequency of the transition involving the levels with energies $E_u$ and $E_l$ (upper and lower, respectively, where $h\nu_{ul}=E_u-E_l$).  From this expression we can derive the maximum optical depth of a transition ($\tau_o$) as well as the peak of the temperature profile ($T_L^o$) of the spectral line.

Under the approximation that the excitation of the molecule is in LTE, the population of all molecular energy levels are determined by an unique temperature ($T_{ex}$). Then one can relate  the column density in the upper level $N_u$ of the transition to the total column density of the molecule $N$ as
\begin{equation}
N_u= \frac{N}{Q(T_{ex})}~g_u~e^{-E_u/kT_{ex}}
\label{eq.LTE}
\end{equation}
where $Q(T_{ex})$ is the temperature dependent partition function of the molecule and $g_u$ and $E_u$ the degeneracy and energy of the upper level of a given transition, respectively. 
Therefore, merging Eq.~\ref{eq.tau} and~\ref{eq.LTE} one can obtain
\begin{equation}
\tau_\circ =\frac{c^3}{8.515736\,\pi\nu_{ul}^3~\Delta {\rm v}} A_{ul}~ \frac{N}{Q(T_{ex})}~g_u~(e^{-E_l/kT_{ex}}-e^{-E_u/kT_{ex}})
\label{eq.tau0}
\end{equation}

It is interesting to mention some relations between the parameters describing the predicted observed profile.
In the optically thin case ($\tau_\circ<<1$), the expected $T_L$ line profile will have a Gaussian shape,
and therefore the area $A$ or integrated intensity of the line can be calculated as
\begin{equation}
\begin{split}
A=\int_{-\infty}^\infty T_L ~\delta {\rm v} = T_L^\circ \sigma \sqrt{2\pi} = 1.064467 ~ T_L^\circ ~ \Delta {\rm v}
\end{split}
\end{equation}


where $T_L^\circ$ is the line temperature at $\tau=\tau_\circ$, i.e. at the center of the line profile. However, in the optically thick case ($\tau_\circ>>1$) the observed profile is broadened 
and the full width half maximum $\Delta {\rm v}^\tau$ can be related to that in optically thin case as 
\begin{equation}
\Delta {\rm v}^\tau = \Delta {\rm v} ~ \sqrt{ \frac{1}{ln2} ~ln \frac{\tau_\circ}{ln(\frac{2}{1+e^{-\tau_\circ}})} }
\label{eq.GaussianTauFWHM}
\end{equation}

and thus the integrated intensity of the opaque line profile is also affected as
\begin{equation}
A^\tau=\Delta {\rm v}^\tau ~ T_L^\tau ~ k(\tau_\circ)
\end{equation}
where $T_L^\tau$ is the intensity at the central velocity corrected by opacity $T_L^\tau= T_L^\circ \bigl( 1-e^{-\tau_\circ} \bigr)$, and $k(\tau_\circ)$ is a opacity dependent proportionality factor which we approximate to be that of the optically thin case.

The input parameter used by SLIM for modelling (Sec.~\ref{sec.syntheticspectra}) and fitting (Sec.~\ref{sec.fitting}) is the actual line width, $\Delta {\rm v}$, which is the meaningful physical parameter. Broadened line profiles are calculated according to the calculated opacity to simulate and fit the observed profiles.
Thus, SLIM considers the effect of optical depth to simulate the observed profiles and, therefore, it provides the actual molecular column density related to the velocity integrated intensity corrected for optical depth effects. 



\subsection{LTE synthetic spectra}
\label{sec.syntheticspectra}
Using the spectroscopic molecular parameters (Section~\ref{sec.database}) and the radiative transfer formalism presented in Sec.~\ref{sec.LTE}, SLIM generates LTE synthetic molecular spectra. The synthetic spectra are generated and controlled through the input physical parameters: 
\begin{itemize}
    \item[\textbullet] column density  ($N$)
    \item[\textbullet] excitation temperature ($T_{ex}$)
    \item[\textbullet] radial velocity of the source ($\rm v_\circ$) 
    \item[\textbullet] full width half maximum of the line ($\Delta \rm v$)
    \item[\textbullet] source size  ($\theta_s$)
\end{itemize}
as well as the corresponding parameters of the continuum defined in (Sect.~\ref{sec.continuum}).
A user friendly GUI interface (Fig.~\ref{fig.SLIMscreenshot}) allows the user to change the individual input parameters using sliders with on-the-fly visualization of the synthetic line profiles superimposed on the observed spectra. The highly interactive interface in SLIM allows for an easy evaluation of the effect of changes in the different physical parameters (e.g., $T_{ex}$) on the LTE spectra. 

To optimize the computing time, the synthetic spectra is only calculated for the range of radial velocities considered to be relevant to fully sample the line profile. The range for the radial where the synthetic spectra is computed is defined by the velocities with line intensities larger than the threshold of $10^{-4}$ relative to its peak intensity. For each Gaussian line profile, as assumed by SLIM (Sect.~\ref{sec.lineprofiles}), this translate into a velocity range of $\lvert$ v-v$_\circ \lvert>$1.75$\Delta$v. Both the plotting and fitting (discussed in Sect.~\ref{sec.fitting}) of the line profiles are also restricted to the data in that range. In the case of simulation of LTE line profile which are unresolved by the spectral resolution of the observations, SLIM calculates the simulated intensity integrated over the observed channel width to make a straightforward comparison with the data.

In the following we provide further details on how the synthetic spectra are calculated.

\subsubsection{Synthetic spectra for multiple molecules with different components}
\label{sec.syntheticmultimolecule}





SLIM produces the LTE synthetic spectra from multiple molecules by considering that the total emission can be described as a linear superposition of the spectra of the individual molecules. In the model, every  molecule can have their own physical parameters. In addition to multiple molecules simulation, SLIM allows also for different components of each molecule. Different components can be differentiated by changes in any of the physical parameters used in the simulation. In this case, SLIM also assumes a linear superposition of the spectra from the different components of each molecules. The results of the LTE synthetic spectrum for multiple molecules, each one with also multiple components can be generalized from Eq.~\ref{eq.TlineradiativeTransfer} as

\begin{equation}
T_L= \sum_m \sum_c \bigl( J_\nu(T_{ex}^{m,c})- J_\nu(T_{c})-J_\nu(T_{bg}) \bigr) ~ (1-e^{-(\sum_t \tau_\nu^t)^{m,c}})
\label{eq.summatorylinetemperature}
\end{equation}
where $m$ and $c$ are the index running for the different molecular species and their components, and $t$ following  Eq.~\ref{eq.sumtau} is the sum over the transitions of each molecular components.

\subsubsection{Linear superposition of components/molecules}
\label{sec.syntheticlinearsuperposition}

\begin{figure*}
\includegraphics[width=1\textwidth]{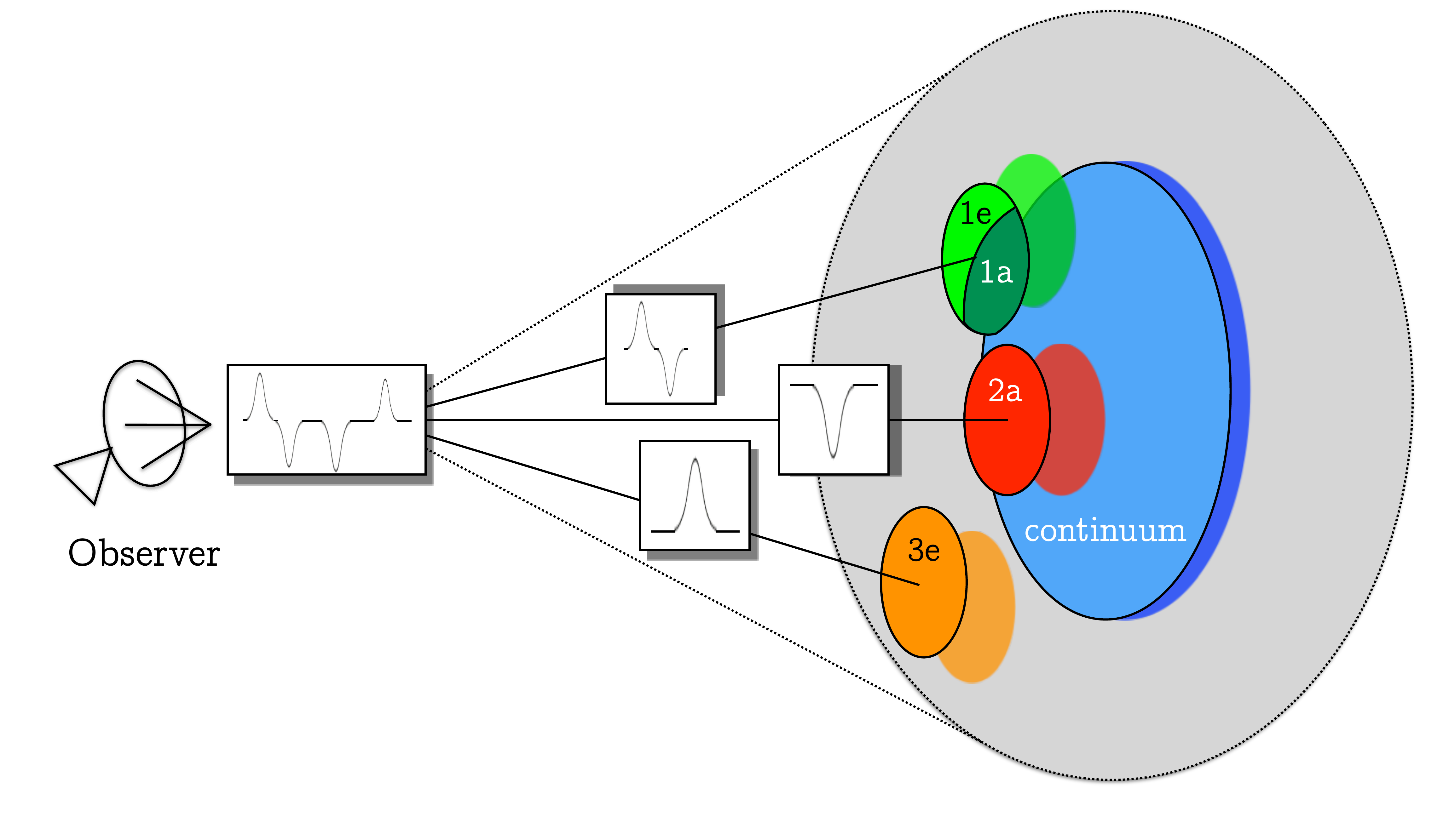}
\caption{Sketch showing an observation of a source consisting of a linear superposition of multiple components. In this scenario, the overall profile shown in left will be the combination of a line in emission (3e), another in absorption against the background source (2a) and a third (1) partially overlapping the background source and therefore observed in both emission (1e) and absorption (1a). See text in Sect.~\ref{sec.syntheticlinearsuperposition} for details.}
\label{fig.sketch}
\end{figure*}

Fig.~\ref{fig.sketch} does illustrates the basic modelling options available in SLIM which can be linearly combined to model spectra of increasing complexity. In Fig.~\ref{fig.sketch} we consider three molecular clouds observed within the telescope primary beam with different sizes, physical properties, kinematics, and continuum coverage. These complex observed emission/absorption line profiles can be easily simulated with SLIM by using different molecular components. 
We briefly discuss how to simulate the molecular clouds along the line of sight with SLIM components using the sketch in Fig.~\ref{fig.sketch} as a guide.:

\begin{itemize}
\item{\textbf{Absorption or emission profiles.} 
By default SLIM considers that molecular components do not have a background continuum source other than the CMB. This is the case of component 3e in Fig.~\ref{fig.sketch}, where line temperature $T_L^{3e}$ will appear in emission as long as $J_\nu(T_{ex}) > J_\nu(T_{bg})$, and will rarely be observed in absorption otherwise (see Sect.~\ref{sec.absorption}).

For each component, SLIM provides the option of considering the radiative transfer effect of a background source, and therefore on the observed line temperatures. That is the case of cloud 2a in Fig.~\ref{fig.sketch}. The properties of the background continuum source are defined as discussed in Sect.~\ref{sec.continuum}. In this case the line will be observed in absorption if $J_\nu(T_{ex}) < (J_\nu(T_{c})+J_\nu(T_{bg}))$. In the case of cloud 2a the cloud only covers a small fraction of the continuum with a covering factor ${\theta_s^{2a}}/{\theta_c} < 1$ in Eq.~\ref{eq.TlineradiativeTransfer}. 
Although each molecule/component can have different different physical parameters and source sizes, the same continuum background spectrum is shared for all components. However, continuum effect is only applied to those molecule/components selected by the user.}

\item{\textbf{Absorption and emission profiles from a single cloud.}} 
SLIM can also simulate the line profiles and derive the physical properties of clouds which are only partially covering the background continuum source, such as the case of component 1 in Fig.~\ref{fig.sketch}. In this case, two SLIM components will be added linearly with their own physical parameters. The solid angles of the two components will be such that $\theta_{1}^2 = \theta_{1e}^2-\theta_{1a}^2$. While 1a will consider the radiative transfer effects of a partially covered continuum emission, 1e will not include such effect. 

\item{\textbf{Kinematics: Velocity structure within the cloud.} 
SLIM allows to simulate the velocity structure of the region of interest by adding multiple velocity components for a given species such as the case of cloud 1 in Fig.~\ref{fig.sketch}.}

\item{\textbf{Non isothermal clouds: Multiple excitation temperatures.}}
So far, we have considered only clouds with uniform excitation temperature, a single $T_{ex}$. As a general case, molecular clouds may show internal thermal structure that can be simulated by the addition of multiple components with different excitation temperatures. SLIM can also handle multiple $T_{ex}$ clouds by creating multiple molecular components with different $T_{ex}$. The SLIM parameters for every component can be selected independently.

However, detecting different temperature components from the observed data is sometime difficult since they usually have similar radial velocities and line widths. Moreover, if the number of transitions available is limited, it may not be straightforward to differentiate between multiple temperature components and the effects of opacity or non-LTE conditions \citep[see Sect.~\ref{sec.rotdiagrams} and][]{Goldsmith1999}. To aid the user in the identification of multiple excitation temperature components, SLIM provides the functionality to generate and plot rotational diagrams (Sect.~\ref{sec.rotdiagrams}) using the observed line intensities. The resulting plots, similar to that shown Fig.~\ref{fig.RotationalDiagram}, can guide the initial estimates of components/temperatures. The user can derive the initial $T_{ex}$ and molecular column densities by fitting straight line(s) to the rotational diagram and use them as the initial parameters for the simulation and/or the fitting of the observed line profiles.

\end{itemize}

The resulting observed line profile (left box in Fig.~\ref{fig.sketch}), will be, according to Eq.~\ref{eq.summatorylinetemperature}, the linear superposition of components and molecules.
Apart from simulating and fitting observed spectra, the user can also generate and save synthetic spectra for any molecule(s) in the catalogs (Sect.~\ref{sec.database}) for a selected frequency range and spectral resolution, where Gaussian noise can also be added to the simulated spectra.  

It is important to note a few details regarding the simulation of spectra:
\begin{itemize}
\item{The sum of optical depths is only considered per component, so the optical depth of the transitions of a given molecular component are added. This is not the case for the linear superposition of components/molecules, where the sum is done in temperature/flux density units, and thus it is assumed that the different components/species are not radiatively coupled. 
While this assumption may be valid for velocity components, which might be actually separated within the resolution element of our observations, it may not be appropriate for multiple temperature components in the line of sight depending on the radial velocity differences and the line widths of the different components.}
\item{When modeling transitions from different observations and observatories, the beam sizes will be different. When generating the synthetic spectrum the beam size of each of the spectra is taken into account to return the appropriate line intensities (Sect.~\ref{sec.observables}).}
\end{itemize}




\subsection{Fitting Algorithm: AUTOFIT}
\label{sec.fitting}





The AUTOFIT function of SLIM performs a non-linear least-squares fitting of simulated spectra to the data to find the optimal free parameters ($N$, $T_{ex}$, $v_{\rm lsr}$, $\Delta \rm v$ and $\theta_s$) used to generate the synthetic LTE spectra. It uses the Levenberg-Marquardt (L-M) algorithm (\citealt{levenberg1944,marquardt1963})
as implemented in the Herschel Common Software System\footnote{The description of the Java implementation can be found in \url{http://herschel.esac.esa.int/hcss-doc-12.0/load/hcss_drm/ia/numeric/toolbox/fit/doc/reference.html#glossary}.} (see \citealt{press2007}). 
The L-M algorithm combines the gradient descent method and the Gauss-Newton method to minimize the $\chi^2$ function:

\begin{equation}
\chi^2 = \sum_{i} \left( \frac{T^{LTE}_i-T^{data}_i}{\sigma_i^2} \right)^2
\end{equation}

where the $T^{LTE}_i$ and  $T^{data}_i$ are the intensities of the LTE simulation and the observed data at the spectral channel $i$, respectively, and $\sigma_i$ is the $rms$ noise of the data when available. When the noise is not available the weight  1/$\sigma^2$ is set to unity. The sum is over the relevant data channels as defined in the previous section for every transition.
The SLIM implementation of the L-M algorithm does not include a priori limits set on the parameters, but they are checked during the iterative fitting procedure. 
Thus, the fitting procedure does an initial check for degeneracies in the input parameters (e.g. guesses of the excitation temperature that are too large compared to the energy levels of the observed transitions, free sources sizes when the result is degenerated because all transitions have been observed with the same telescope/beam, ...). Additionally, every ten iterations of the L-M algorithm, the validity of the of the fitted parameters is also checked
(e.g., negative excitation and molecular column densities, too large excitation temperature compared with those of the observed transition energy levels or fitted source sizes too small compared with the beam)
In case of failing the validation, the fitting procedure stops and requests an action from the user to help the convergence of the fit by changing the input parameters.

In addition to the fitted value, 
AUTOFIT also provides their associated error and a measure of the goodness of the fit. The errors in the fitted parameters are the 1$\sigma$ standard deviations derived as follows:	 

\begin{equation}
   \sigma_{p_i} = \sqrt{\matr{H^{-1}(i,i)} \frac{\chi^2}{(N-K)}}
\end{equation}  
with i = 1....K is the number of fitted parameters, N is the number of spectral data channels used in the fit, and H is the Hessian matrix derived from the inner product of the transpose of the Design matrix  D(i,j) by itself.

\begin{equation} 
   \matr{H} = \matr{D^T} \times \matr{D} 
\end{equation}  

The Design matrix is built from the partial derivatives of the LTE line profile model to each of its parameters at the velocity of every spectral channel: 

\begin{equation} 
\matr{D(i,j)} = \left(\frac{\partial\ T^{\rm LTE}}
              {\partial q_j}\right)_i
\end{equation}

where $i$ refers to the channel $i$ and $j$ to the parameter to be fitted. For completeness, in Appendix ~\ref{sec.derivatives} we present the analytic derivative equations used to generate the Design matrix. The estimated errors in the fitted parameters take into the account all the data channels used in the fit, and therefore they decrease when the number of the relevant data channels, $N$, increases roughly with a factor of $\sqrt{N}$ .

In addition to the error of the parameters, SLIM also provides information on the goodness-of-fit through the reduced/normalized $\overline{\chi}$ defined as: 

\begin{equation}
\overline{\chi} = \sqrt{\frac{\chi^2}{(N-K)}}
\end{equation}

$\overline{\chi}$ does not depend on the number of data channels used in the fit as the errors of the parameters. The value of $\overline{\chi}$ will be different depending on the weighting of the data used in the fit. In the case that the 1/$\sigma^2$ weight is used, $\overline{\chi}$ should be close to 1, that is the same order of the $rms$ noise of the data, indicating a good quality fit.  Large values of $\overline{\chi}>>$ 1, indicate a poor fit and small values of $\overline{\chi}<$1 suggest over fitting, i.e. the model is likely fitting some of the data noise. If the rms noise of the data in not provided to AUTOFIT, i.e weight = 1 for all data points, the $\overline{\chi}$ has to be derived as the ratio between the $\overline{\chi}$ provided by AUTOFIT and the estimated rms noise of the data. Caution should be taken with the estimated $\overline{\chi}$ when the number of data channels is only slightly larger than number of fitted parameters since it depends of N-K. In this case the $\overline{\chi}$ might be meaningless. 

Since AUTOFIT uses spectroscopically-resolved line profiles, this function cannot be applied for data that do not resolve the line profiles. Upper limits to the molecular column densities can also be calculated automatically when the lines are not detected. SLIM estimates the local rms noise of a spectroscopic channel, $\sigma$, from the data by fitting a baseline to a selected region, free from line emission, of the spectrum displayed in the plot. The upper limit to the column density is then calculated from the upper limit to the integrated intensity

\begin{equation}
\int T_R \Delta {\rm v} > 3\sigma\sqrt{\frac{\Delta {\rm v}}{\delta {\rm v}}}
\end{equation}

where $\Delta {\rm v}$ and  $\delta {\rm v}$ are the SLIM linewidth parameter and the spectral resolution of the data respectively. The upper limit of the integrated intensity is converted to the upper limit to the column densities using the SLIM parameters, $T_{ex}$ and $\theta_s$.






\section{Discussion: Usual molecular line radiative transfer approximations in the literature. Limitations and degeneracies}
\label{sec.approximations}

For the sake of simplicity, a number of assumptions and/or approximations are commonly used in the literature to derive physical parameters from observed spectra \citep[see][]{Mangum2015}.
In their paper, they explore the optically thin, thick and R-J regimes. 
However, with tools like MADCUBA-SLIM, it becomes unnecessary to rely on such approximations.
Rather than exploring the radiative transfer expression for different limiting cases  
as done by \citet{Mangum2015}, in this section we do explore the most usual approximations used in the literature for column density determination. Here we include the use of rotational diagrams, the analysis of absorption lines assuming a strong background continuum source, and the opacity/column density calculation from hyperfine splitting line intensity ratios. With SLIM, not only the use of such approximations is not required, but it also offers the possibility to explore the uncertainties and the range of validity of those approximations. The figures in this section, generated from the output models of SLIM, aim to explore the errors resulting from the use of approximations in the ranges where they break.



\subsection{Rotational diagrams}
\label{sec.rotdiagrams}
Rotational diagrams are one of the most used approximations for deriving column densities and excitation temperatures. The basics of these approximation are described in numerous references \citep[i.e.][]{Goldsmith1999,Mart'in2006}.
For this approximation the basic assumptions are:
\begin{itemize}
    \item[\textbullet] {No CMB: $T_{bg}=0$}
    \item[\textbullet] {No background continuum source: $T_{c}=0$}
    \item[\textbullet] {Optically thin emission: $\tau_\nu<<1$}
\end{itemize}

Thus Eq.~\ref{eq.TlineradiativeTransfer} can be combined with Eqs.~\ref{eq.tau} and ~\ref{eq.LTE}, and in logarithmic we obtain the linear expression

\begin{equation}
    log \left(\frac{8\pi k \nu^2}{h c^3 g_u}\int T_L\delta {\rm v}\right) = log (\frac{N}{Q(T_{ex})}) - \frac{E_u}{k T_{ex}} log(e)
\label{eq.rotationaldiagram}
\end{equation}

The first two assumptions will hold as long as $J(T_{ex})>>J(T_c)+J(T_{bg})$ which will be generally true for sources with no background continuum source. However, for low excitation $T_{ex}\sim T_{bg}$ the effect of the CMB background will not be negligible. This may be even more relevant for high redshift studies where the the CMB temperature scales with redshift as $T_{bg}= 2.73~(1+z)$. At $z=7$ the cosmic microwave background temperature is $\sim$22~K, and will therefore play an important role in the excitation of the molecular clouds.

\begin{figure}
\includegraphics[width=0.5\textwidth]{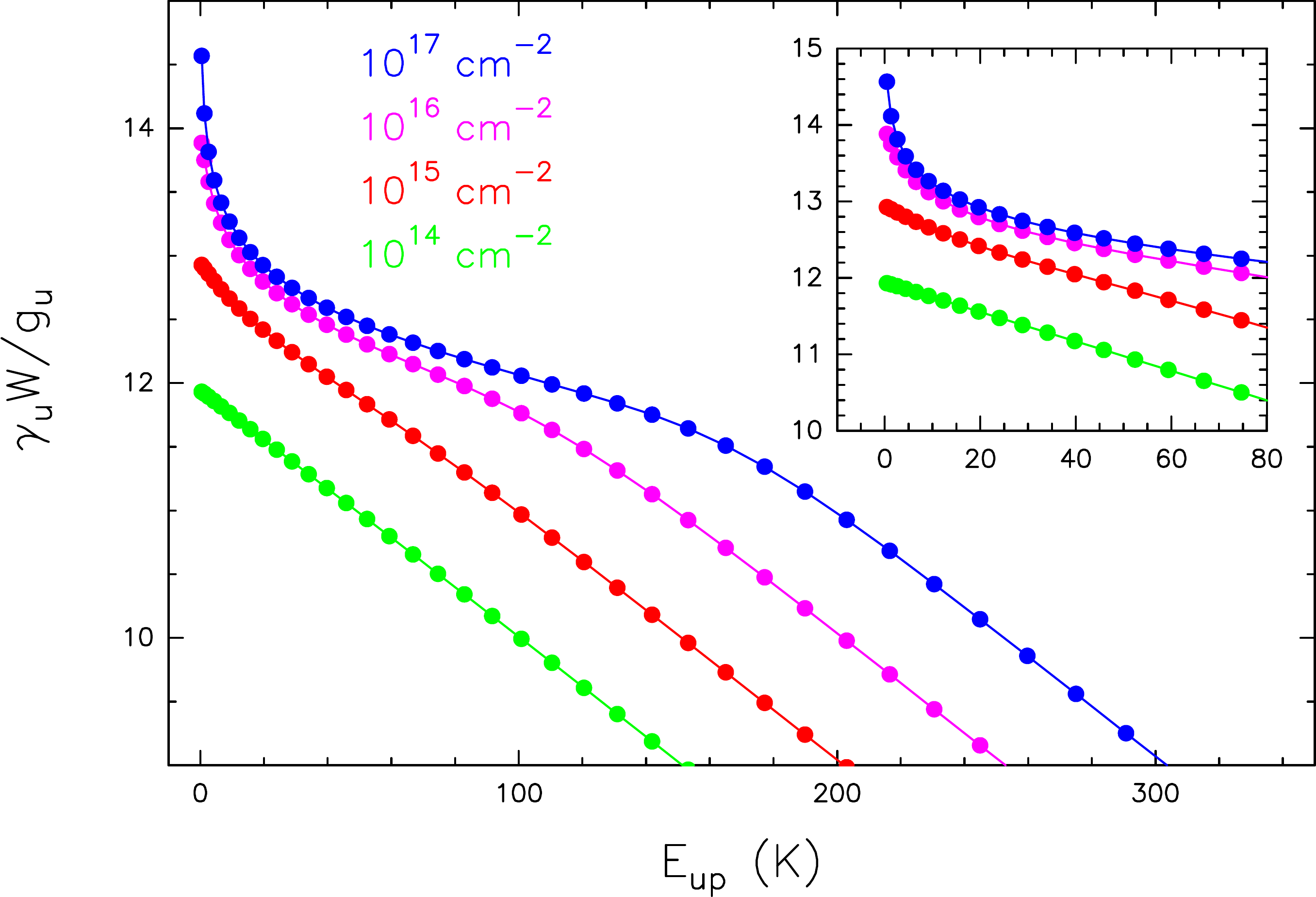}
\vskip5mm
\includegraphics[width=0.5\textwidth]{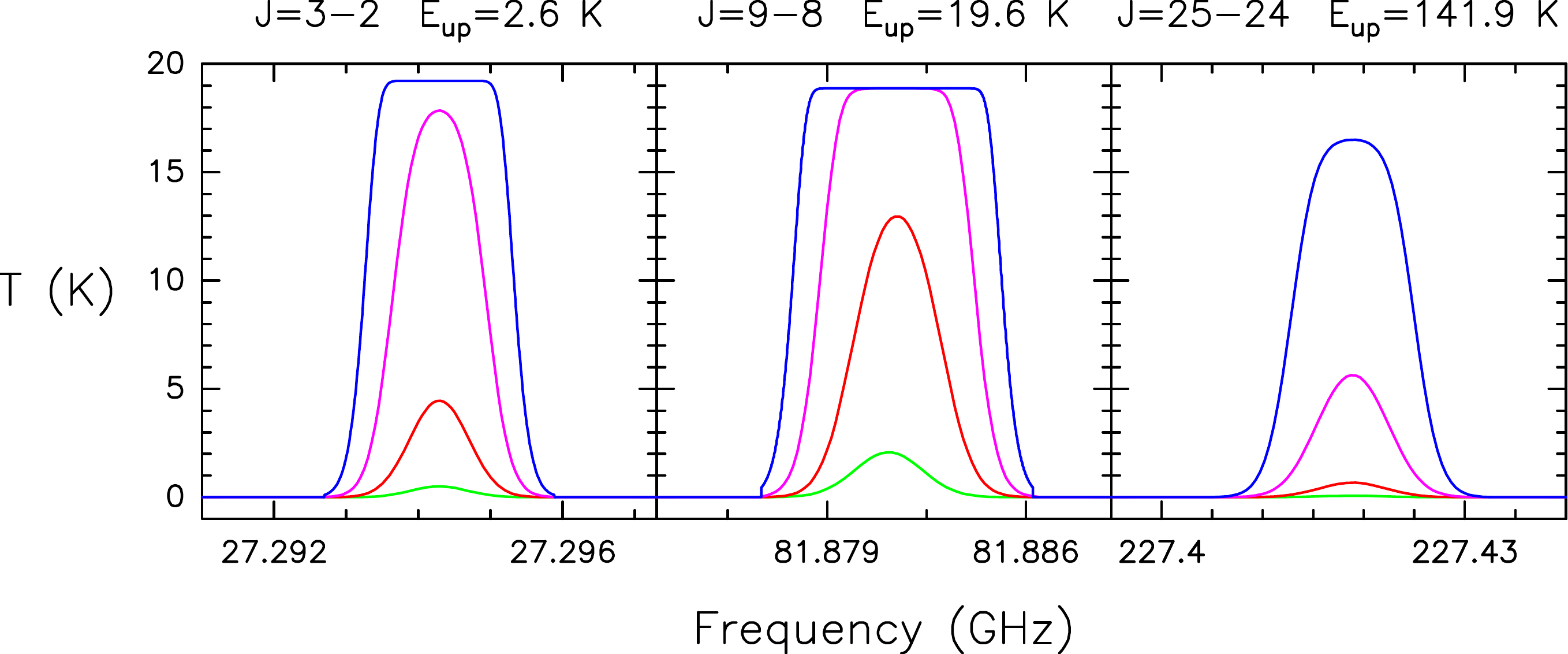}
\caption{(Upper panel) Rotational diagram of the first 36 rotational transitions of HC$_3$N generated from the SLIM output synthetic spectra. Physical parameters of T$_{\rm ex}$=22 K, $\Delta$= 10 km s$^{-1}$ and extended source size have been used. The inset figure on the top left corner matches the range of energies displayed in Fig.~2 from \citet{Goldsmith1999}.
(Lower panel) Synthetic spectra of three transitions at representative upper level energies for the four column densities used to generate the rotational diagram. The effect of opacity is clearly seen in both line saturation and broadening.
}
\label{fig.RotationalDiagram}
\end{figure}

Leaving aside the considerations on background temperature, the rotational diagram approximation will break for moderate to high column densities where significant deviations from linearity will occur due to the effects of increasing optical depth, which will be transition dependent.
This is illustrated in Fig.~\ref{fig.RotationalDiagram} where the HC$_3$N rotational diagram has been generated out of the integrated intensities simulated by SLIM for molecular column densities ranging from $10^{14}$ to $10^{17} \rm cm^{-2}$. We simulated the emission of HC$_3$N assuming $T_{ex}=22~K$ and $\Delta {\rm v}=10~\rm km~s^{-1}$ . The species and the physical parameters were selected to match the diagram shown in Fig.~2 from \citet{Goldsmith1999}. The inset diagram in the top left corner of Fig.~\ref{fig.RotationalDiagram} does actually match the Figure in \citet{Goldsmith1999}, with the main difference of their Figure y-axis being in natural logarithm scale and slightly different column densities.
The line profiles for three out of the 36 rotational transitions used in the diagram are shown for a wide range of upper energy levels, where the effect of line saturation and line broadening (Eq.~\ref{eq.GaussianTauFWHM}) are evidenced even for high energy transitions.

The curvature of the rotational diagram (blue and magenta points in Fig.~\ref{fig.RotationalDiagram}), together with an insufficient sampling in the number of observed transitions, can lead to the miss-identification of multiple temperature components. This section shows that SLIM correctly takes into account the effects of increasing optical depth in the molecular transitions. 
In any case, as mentioned in Sect.~\ref{sec.syntheticlinearsuperposition} on non-isothermal modeling, the use of rotational diagrams may help as a first guide for initial fitting parameters in models with multiple temperatures.

It is important to mention here the limitation regarding the LTE assumption. Fig.~6 in \citet{Goldsmith1999} shows how non-LTE effects do also break the linearity when the emission is not thermalized. Thus, if molecular hydrogen volume density is below the critical density ($n(H_2)<n_{crit}$) of any given transition, the LTE assumption in SLIM will also break. As a result, curved rotational diagrams would end up in a degeneracy between low density sub-thermal excitation or multiple high density components such as that observed in the CO ladder towards Sgr~A$^*$ \citep{Goicoechea2015}.

\subsection{Absorption profiles}
\label{sec.absorption}

Most studies of absorption molecular profiles do display the normalized spectra in order to directly measure the fraction of the continuum absorbed by foreground gas. Thus, instead of defining the line temperature after subtracting the continuum in Eq.~\ref{eq.TlineradiativeTransfer}, the fraction of absorption of the continuum can be written as


\begin{figure}
\includegraphics[width=0.45\textwidth]{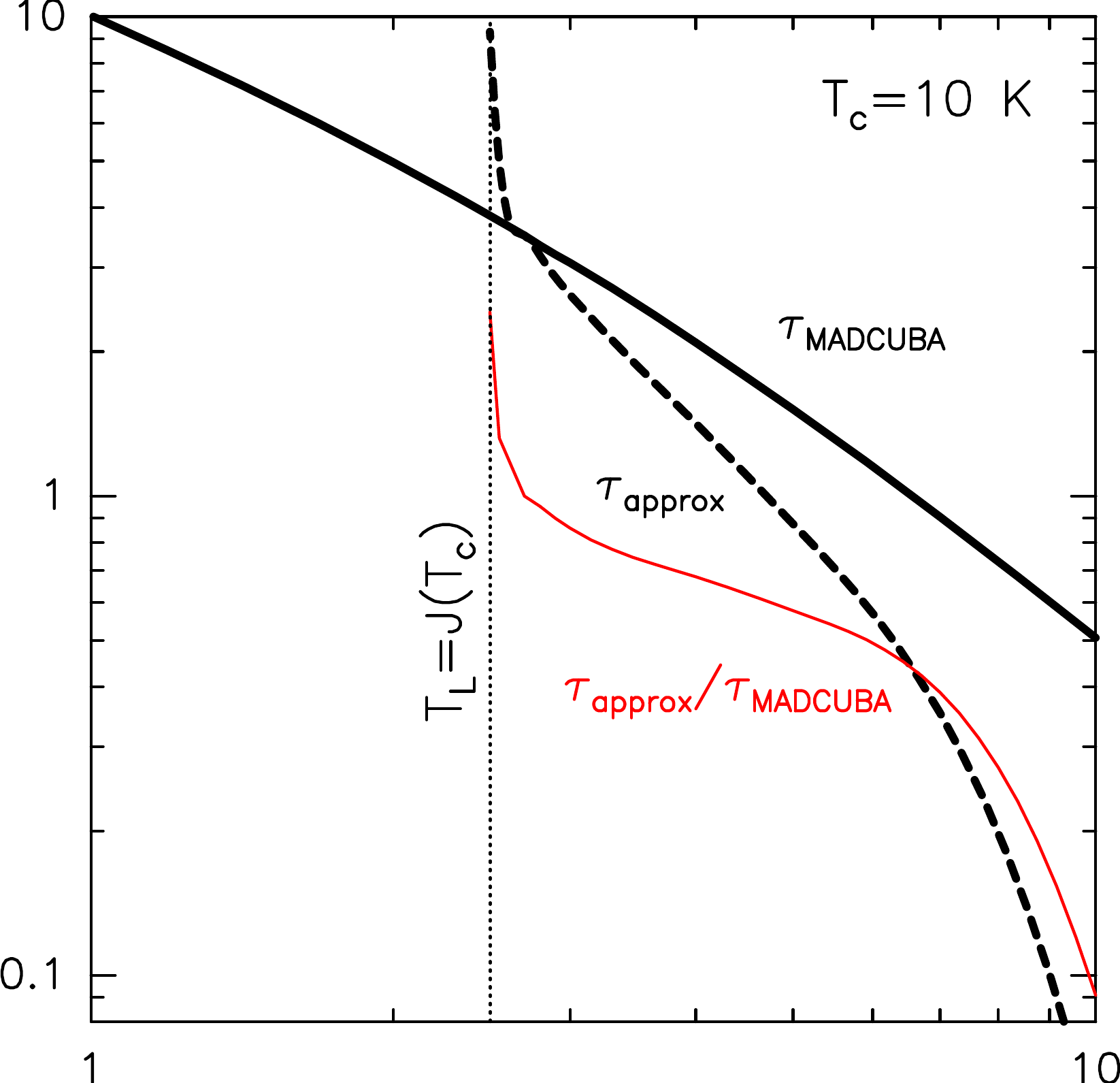}
\vskip5mm
\includegraphics[width=0.45\textwidth]{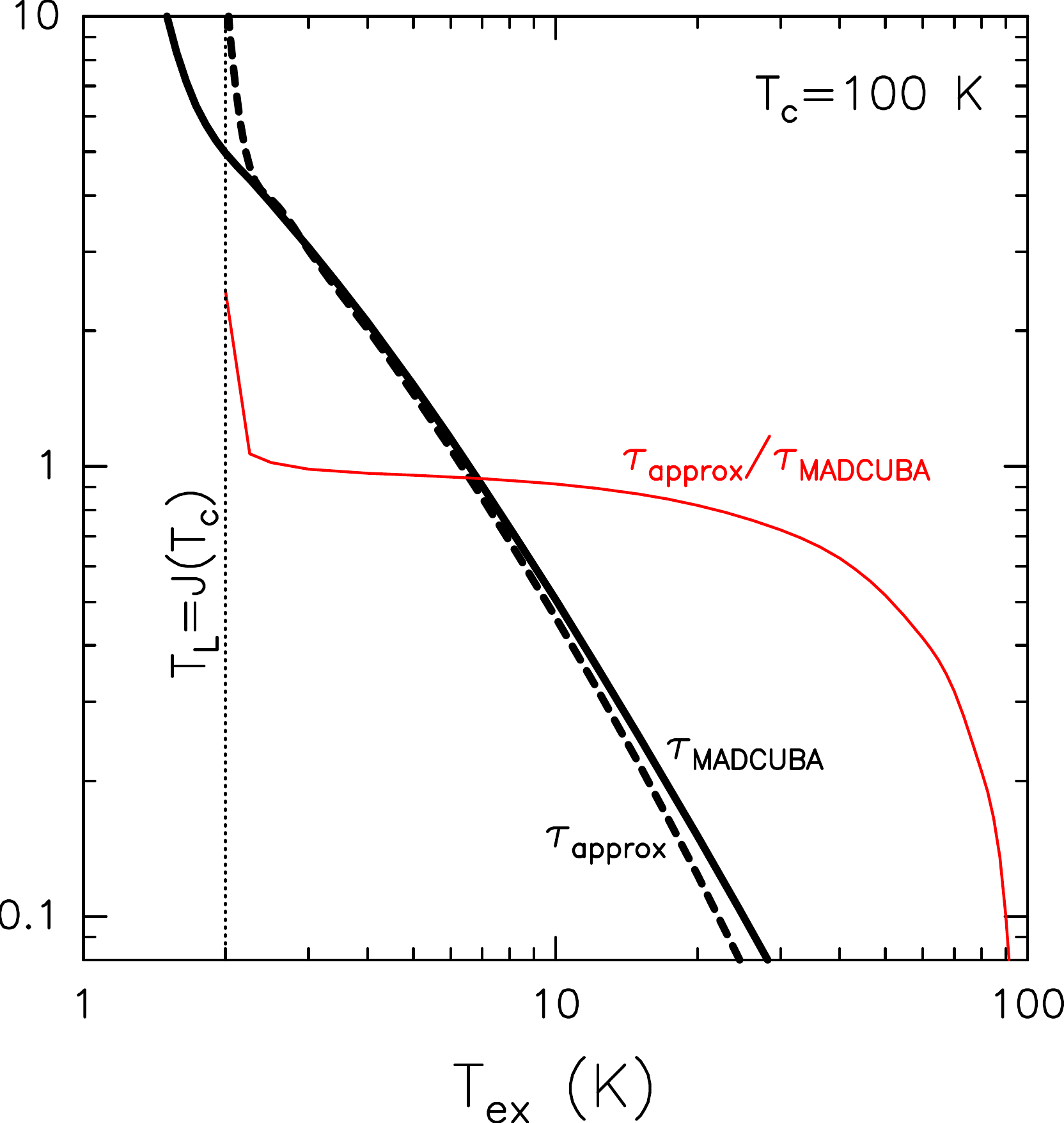}
\caption{Optical depth of the absorption profile of CO $1-0$ calculated by MADCUBA (black continuous line) for $\Delta$v=10 km s$^{-1}$, log~$N$=17 cm$^{-2}$ and $T_{bg}=2.73$~K. From the values measured in the synthetic spectra and using Eq.~\ref{eq.absorptionapprox}, the approximate optical depth is calculated (black dashed line). The ratio between the two optical depth estimates is shown in red line. Top and bottom panels are equal but for the continuum temperatures of 10 and 100~K, respectively. Vertical dotted line represent the temperature for which the line does absorb completely the continuum emission (i.e. $T_L=J(T_c)$), and beyond which the CMB continuum is absorbed.}
\label{fig.absorption}
\end{figure}

\begin{equation}
    \frac{T_{L}}{J(T_{c})} = \frac{\bigl( J_\nu(T_{ex})-c_s~J_\nu(T_{c})-J_\nu(T_{bg}) \bigr) (1-e^{-\tau_\nu})}{J(T_c)}
    \label{eq.absorption}    
\end{equation}

If we assume that the absorption is occurring against a very bright continuum source, which implies that
\begin{itemize}
    \item[\textbullet]{$T_c >> T_{bg}$}
    \item[\textbullet]{$T_c >> T_{ex}$}
\end{itemize}
then Eq.~\ref{eq.absorption} reduces to
\begin{equation}
    \tau_\nu=-ln\Bigl(1-\frac{T_{L}}{c_s~J(T_{c})}\Bigr)    
    \label{eq.absorptionapprox}    
\end{equation}
commonly used to quickly estimate the opacity of the absorption profiles towards bright continuum emitters \citep[i.e.][]{Muller2011}.
This expression, where usually $c_s$ is considered to be unity, is independent of the excitation temperature of the gas. Fig.~\ref{fig.absorption} shows the optical depth of the CO $J=1-0$ profile calculated by MADCUBA for fixed values of $\Delta$v=10 km s$^{-1}$, log~$N$=17 cm$^{-2}$ and $T_{bg}=2.73$~K, as a function of varying $T_{ex}$. Fig.~\ref{fig.absorption} also displays the optical depth calculated based on the generated synthetic spectra and with the approximation in Eq.~\ref{eq.absorptionapprox}. 
The optical depth in Fig.~\ref{fig.absorption} directly calculated by MADCUBA is shown as a black thick solid line, and the one estimated from the synthetic absorption profile using Eq.~\ref{eq.absorptionapprox} as a black dashed line. These optical depths are calculated as a function of the assumed $T_{ex}$ and for two background continuum black bodies with temperatures $T_c$ of 10 and 100~K (top and bottom panels in Fig.~\ref{fig.absorption}, respectively).

Fig.~\ref{fig.absorption} shows that the optical depth is actually temperature dependant. However, it also illustrates, as explained below, the range in which this approximation to estimate the optical depth is valid, which is relative to the brightness of the illuminating continuum source.
The ratio between both optical depths is also displayed as a red line to better show the region where the approximation holds.

Fig.~\ref{fig.absorption} shows that when $T_{ex}$ approaches the CMB temperature (in this case $T_{bg}$=2.73~K), the approximation in Eq.~\ref{eq.absorptionapprox} goes to infinity since the continuum is completely absorbed ($T_L=J(T_c)$, dotted vertical line in Fig.~\ref{fig.absorption}). Although it is an unusual situation, the approximation in Eq.~\ref{eq.absorptionapprox} will not apply for temperatures below the CMB temperature ($T_{ex}<T_{bg}$), where the normalized absorption gets negative since it does absorb the CMB continuum \citep[see][and references therein]{Martin-Pintado1985}. Similarly, high optical depths can be the result of very high column densities. In this situation, the fit to the line profile by MADCUBA and and in particular line profile broadening (Eq.~\ref{eq.GaussianTauFWHM}) will allow for a more accurate estimate of the optical depth than just based on the intensity of the absorption.

For excitation temperatures approaching that of the background continuum ($T_{ex}\sim T_c$), the absorption will vanish and therefore the opacity calculated based on the approximation in Eq.~\ref{eq.absorptionapprox} will be severely underestimated. This is particularly critical for faint background illuminating continuum sources \citep[i.e. search for absorption against faint GRB events;][]{deUgartePostigo2018} where a non detection of absorption may be either low absorbing column density or significant amounts of gas not seen in absorption due to $T_{ex}\sim T_c$.

For temperatures in between the CMB temperature and the temperature of the background continuum source, the optical depth estimated with Eq.~\ref{eq.absorptionapprox} has a relatively narrow range of validity out of which the approximation, and therefore the derived column density, will be over/underestimated for low/high excitation temperatures, respectively. 

With the possibility of using SLIM the user does not need to make the assumptions mentioned above, and through educated guesses of the excitation temperature it will be possible for the user to estimate the errors in the optical depths and column densities derived from the observations. 
Moreover, the SLIM simulation will also help the user to prepare the observations that will provide the best estimate of the excitation temperature by measuring several transitions.

\begin{figure*}
\includegraphics[width=1\textwidth]{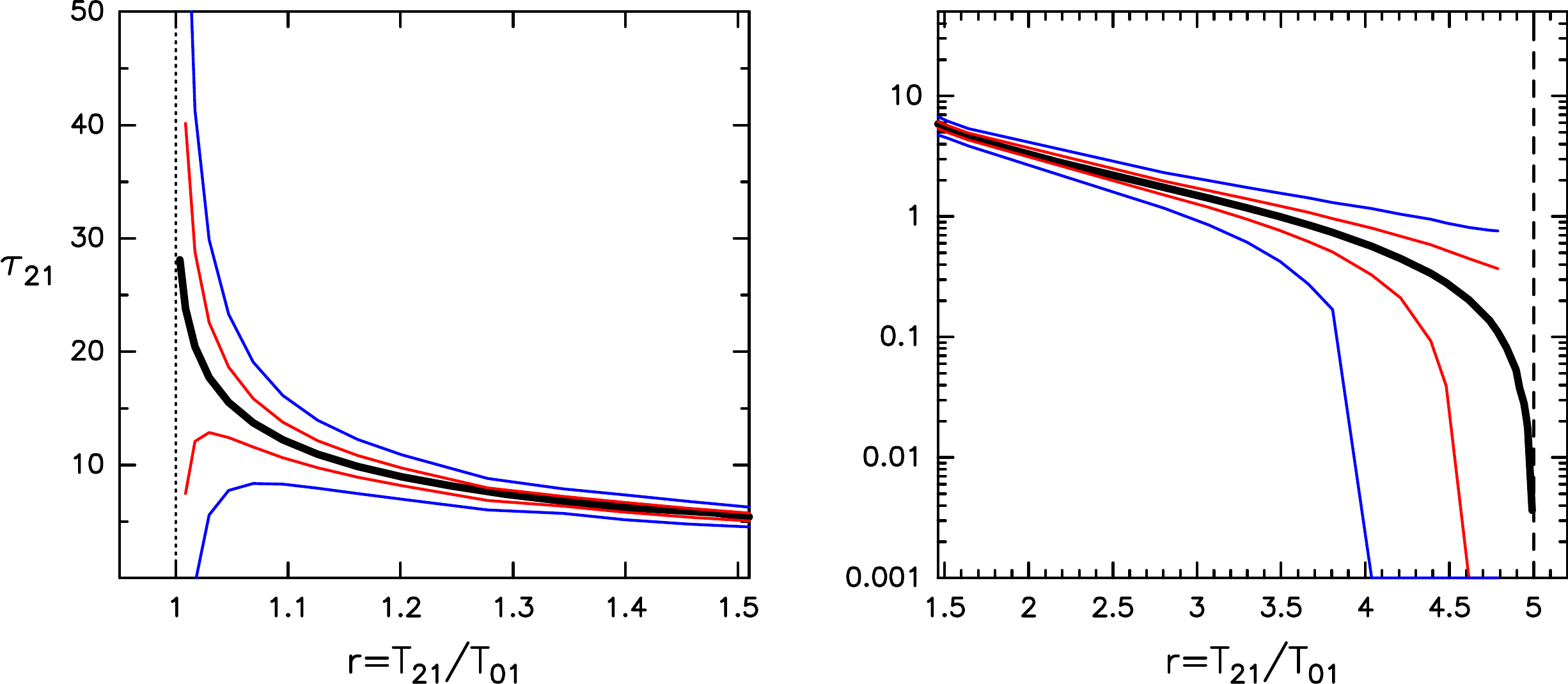}
\caption{Optical depth to the $F=2-1$ hyperfine transition of N$_2$H$^+$ as a function of ratio of the line intensities of the $F=2-1$ to the $F=0-1$ shown in black. The line width used in the SLIM simulations is $\Delta$v= 2 km s$^{-1}$ to avoid blending between hyperfine transitions. In red and blue we represent the propagated uncertainties considering that the $F=2-1$ hyperfine transitions of N$_2$H$^{+}$ is detected with SNR of 50 and 20, respectively. See Sect.~\ref{sec.hyperfine} for details.}
\label{fig.opacity}
\end{figure*}

\subsection{Molecular line opacities from hyperfine splitting}
\label{sec.hyperfine}
As commonly used in the literature \citep[i.e.]{Henkel1998}, the ratio of the line intensities of two molecular transitions (or blended groups of transitions) can be written as

\begin{equation}
\frac{T_{L,1}}{T_{L,2}}=\frac{1-e^{-\tau_1}}{1-e^{-\tau_2}}
\label{eq.hyperfinelineratio}
\end{equation}
where this expression implicitly assumes
\begin{itemize}
    \item[\textbullet]{No background continuum source: $T_{c}=0$}
    \item[\textbullet]{LTE conditions: $T_{ex,1}=T_{ex,2}$}
\end{itemize}

Hyperfine transition splitting in molecules provides a way of directly infer the optical depth of the emission through the observed hyperfine line intensity ratios. In the optically thin regime, the ratio between any given pair of hyperfine lines is directly related to the ratio between their optical depths $\tau_1/\tau_2$, under the above mentioned assumptions. Since most spectroscopic parameters (frequecies, energy levels, and einstein coefficients) are almost identical, from Eq.~\ref{eq.tau0} this optical depth ratio will be equal to the ratio between involved hyperfine transition upper level degeneracies $\tau_1/\tau_2=g_{u,1}/g_{u,2}$.

As a case study we consider here the $J=1-0$ transition of N$_2$H$^+$ which is splitted into three hyperfine components $F=2-1$, $1-1$, and $0-1$ at 93173.7, 93171.88, and 93176.13~MHz, respectively (based on the spectroscopy from the JPL catalog).
In Fig.~\ref{fig.opacity} we show the opacity of the brightest component $\tau_{21}$ as a function of the ratio of the $F=2-1$ to the $F=0-1$ transition. In the optically thin regime, as explained above, the ratio between the line intensities will approach that of the upper level degenacies ratio ($g_{21}=15$ and $g_{01}=3$), and therefore $\tau_{21}/\tau_{01}=5$. On the other hand, as the optical depth grows towards the optically thick regime, both spectral line will saturate and their ratio will asymptotically approach unity (see Eq.~\ref{eq.hyperfinelineratio}).

Despite the relation between line intensities and optical depths, albeit the approximations above, the intrinsic uncertainties to the line temperature measurements may translate into large uncertainties in the estimated optical depth as we get close to the limiting cases. 
In Fig.~\ref{fig.opacity}, we show the actual relation with a thick black line, while the red and blue lines show the uncertainty in the derived optical depth by assuming a signal-to-noise ratio in the $F=2-1$ transition ($SNR=T_{\rm 21}/rms$, where $rms$ is the noise of the spectra) of 50 and 20, respectively.
The uncertainty in the optical depth is calculated via the error propagation (using its numerical derivative) as 
$\Delta \tau_{\rm 21}=\left(\frac{\partial \tau_{\rm 21}}{\partial r}\right)\Delta r$, where the uncertainty on the line ratio ($r$) is calculated as $\Delta r=r\sqrt{1+r^2}/SNR$.
In the optically thin regime, as we approach the theoretical value of 5, the uncertainties in the optical depth, and therefore in the column density determination may vary by several orders of magnitude.
Furthermore, towards the optically thick regime, though not so severe, the optical depth can be easily uncertain by a factor of a few. It is only within the range of $\tau_{21}$ between $\sim 1-10$, that the optical depth would be well constrained.

We note that the large uncertainties in the optically thin regime is a result of the decreased signal-to-noise ratio of the $F=0-1$ line, which being a factor of 5 fainter than the $F=2-1$, would be detected at a signal-to-noise ratio of 10 and 4, for the two examples shown in Fig.~\ref{fig.opacity}.
The same study could be carried out with the ratio of the $F=2-1$ to $1-1$, since the later is brighter then the $0-1$ with $\tau_{21}/\tau_{11}=g_{21}/g_{11}\sim1.66$. However, in this case, the dynamic range of the ratio will be limited to the range from 1.66 in the optically thin regime to 1 in the thick end.

\begin{figure}
\includegraphics[width=0.45\textwidth]{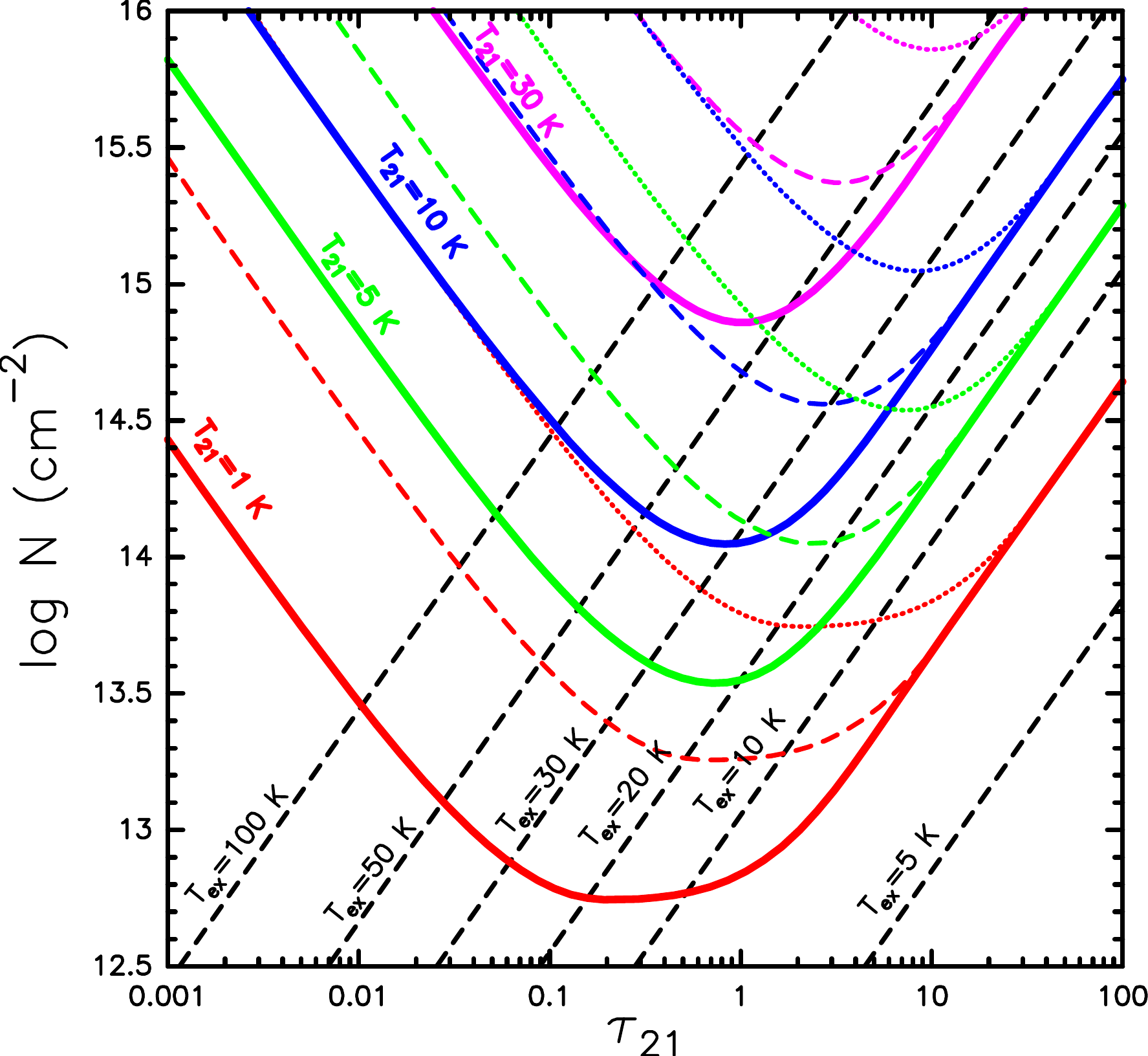}
\caption{Column density of N$_2$H$^+$ as a function of the estimated optical depth (Fig.~\ref{fig.opacity}) of its $J=1-0, F=2-1$ transition for different assumed excitation temperatures (black dashed lines). In colour we represent isotemperature curves of simulated line intensities in main beam temperature scale for the $26.4''$ primary beam of the IRAM~30~m at the frequency of this transition. The curves are calculated for different source sizes: extended (continous line); 2/3 of the primary beam (dashed); and 1/3 of the primary beam (dotted). The simulation considers a line width $\Delta$v= 2 km s$^{-1}$, and a $T_{bg}=2.73$~K.}
\label{fig.tautocolumndensity}
\end{figure}

Once the opacity is estimated from the line intensity ratio, column density can be derived via an assumption on the excitation temperature as represented with the black dashed lines in Fig.~\ref{fig.tautocolumndensity}. In this Figure, we added the predicted line intensity in main beam brightness temperature scale as it would be observed with the $26.4''$ beam of the IRAM~30m telescope at the frequency of this transition. Each colour represent an iso-line-temperature contour for an extended source (continuous line), and a source size of $2/3$ and $1/3$ of the telescope beam (dashed and dotted line, respectively). This Figure does include the effect of the CMB, but no other background continuum source.

The main idea behind Fig.~\ref{fig.tautocolumndensity} is to illustrate the fact that apart from the line intensity ratio used to derive the optical depth of the emission, the absolute measured line intensity carries out valuable information that may allow to break the degeneracy between optical depth, excitation temperature, and source size (see Eq.~\ref{eq.unitssizes}).

In other words, once the opacity has been measured with the hyperfine structure, it is not possible to determine the column density with prior assumptions on the excitation temperature and source size, since this may not fit the actual line intensity measured. Since SLIM fits directly the observed spectrum by varying the physical parameters, both the absolute line intensity and intensity ratio are fitted simultaneously. 

Thus if we have a good a priori information on the excitation temperature, the line intensity may provide constrains on the source size. On the other hand the line intensity can also set constrains to the possible excitation temperatures with a priori information on the source size. In fact, this allows for determination of excitation temperatures based on a single hyperfine splitted transition if not in the optically thin/thick regime. Of course, Fig.~\ref{fig.tautocolumndensity} is valid for this particular transition of N$_2$H$^+$, but similar plots could be generated for other transitions/molecules. Additionally, when a particular transition gets optically thick, the line broadening by opacity (Eq.~\ref{eq.GaussianTauFWHM}) can be used to further break the degeneracy between physical parameters.

The above discussion also applies to not hyperfine splitted transitions. In this case, opacity will not be constrained by the line intensity ratio and the degeneracy will be directly on column density, excitation temperature, and source size. Any combination of these parameters will have to reproduce the observed intensity.

Fig.~\ref{fig.tautocolumndensity} show how by fitting the line profile as a whole with SLIM we can constraint the physical parameters of column density and excitation temperature. More importantly, SLIM allows the user to easily explore the parameter space and observe on-the-fly the effect of the changes in the line profile by varying input parameters.

\section{Conclusion}
In this paper we present the main functionalities of the highly interactive spectroscopic data handling package MADCUBA and its spectroscopy analysis module SLIM. The basic radiative transfer formalism and fitting algorithm used by SLIM have been described in detail.
We discussed the usual approximations used in the literature for different limit cases that simplify the radiative transfer equation in order to derive physical parameters. We show that these approximations may result in both significant deviations in the derived parameters, when the assumptions for such approximations break, as well as in an underestimation of the errors associated to those approximations. More importantly, we show that with modern tools like MADCUBA and SLIM, there is no need to resource to those approximations since we can easily model the observed spectra using the full radiative transfer equation.
We also note that, by design, SLIM, works under the LTE assumption, and therefore it will not apply to non-LTE conditions or under non-LTE excitation.

MADCUBA and SLIM, as other of the self tools open to the community, does allow for an efficient handling of large spectroscopic dataset coming for state-of-the-art radio, millimeter and sub-millimeter facilities.


\begin{acknowledgements}
The MADCUBA and SLIM development has been partially funded through the Spanish grants ESP2013-21697-C05-01, ESP2015-65597-C4-01-R and ESP2017-86582-C4-01-R. 
V.M.R. has received funding from the European Union's Horizon 2020 research and innovation programme under the Marie Sk\l{}odowska-Curie grant agreement No 664931. 
S.M. and V.M.R. acknowledge support from the Joint ALMA Observatory Visitor Program.

\end{acknowledgements}

\bibliographystyle{aa}
\bibliography{SLIMMADCUBA}

\begin{thebibliography}{52}
\expandafter\ifx\csname natexlab\endcsname\relax\def\natexlab#1{#1}\fi

\bibitem[{{Aladro} {et~al.}(2015){Aladro}, {Mart{\'{\i}}n}, {Riquelme},
  {Henkel}, {Mauersberger}, {Mart{\'{\i}}n-Pintado}, {Wei{\ss}}, {Lefevre},
  {Kramer}, {Requena-Torres}, \& {Armijos-Abenda{\~n}o}}]{aladro2015}
{Aladro}, R., {Mart{\'{\i}}n}, S., {Riquelme}, D., {et~al.} 2015, \aap, 579,
  A101

\bibitem[{{Armijos-Abenda{\~n}o} {et~al.}(2018){Armijos-Abenda{\~n}o},
  {L{\'o}pez}, {Mart{\'{\i}}n-Pintado}, {B{\'a}ez-Rubio}, {Aravena},
  {Requena-Torres}, {Mart{\'{\i}}n}, {Llerena}, {Ald{\'a}s}, {Logan}, \&
  {Rodr{\'{\i}}guez-Franco}}]{Armijos-Abendano2018}
{Armijos-Abenda{\~n}o}, J., {L{\'o}pez}, E., {Mart{\'{\i}}n-Pintado}, J.,
  {et~al.} 2018, \mnras, 476, 2446

\bibitem[{{Baez}(2014)}]{Baez2014}
{Baez}, A. 2014, PhD thesis, Universidad Complutense de Madrid

\bibitem[{{Beltr{\'a}n} {et~al.}(2018){Beltr{\'a}n}, {Cesaroni}, {Rivilla},
  {S{\'a}nchez-Monge}, {Moscadelli}, {Ahmadi}, {Allen}, {Beuther}, {Etoka},
  {Galli}, {Galv{\'a}n-Madrid}, {Goddi}, {Johnston}, {Klaassen},
  {K{\"o}lligan}, {Kuiper}, {Kumar}, {Maud}, {Mottram}, {Peters}, {Schilke},
  {Testi}, {van der Tak}, \& {Walmsley}}]{Beltran2018}
{Beltr{\'a}n}, M.~T., {Cesaroni}, R., {Rivilla}, V.~M., {et~al.} 2018, \aap,
  615, A141

\bibitem[{{Colzi} {et~al.}(2018){Colzi}, {Fontani}, {Rivilla},
  {S{\'a}nchez-Monge}, {Testi}, {Beltr{\'a}n}, \& {Caselli}}]{Colzi2018}
{Colzi}, L., {Fontani}, F., {Rivilla}, V.~M., {et~al.} 2018, \mnras, 478, 3693

\bibitem[{{Cosentino} {et~al.}(2018){Cosentino}, {Jim{\'e}nez-Serra},
  {Henshaw}, {Caselli}, {Viti}, {Barnes}, {Fontani}, {Tan}, \&
  {Pon}}]{Cosentino2018}
{Cosentino}, G., {Jim{\'e}nez-Serra}, I., {Henshaw}, J.~D., {et~al.} 2018,
  \mnras, 474, 3760

\bibitem[{{de Ugarte Postigo} {et~al.}(2018){de Ugarte Postigo}, {Th{\"o}ne},
  {Bolmer}, {Schulze}, {Mart{\'{\i}}n}, {Kann}, {D'Elia}, {Selsing},
  {Martin-Carrillo}, {Perley}, {Kim}, {Izzo}, {S{\'a}nchez-Ram{\'{\i}}rez},
  {Guidorzi}, {Klotz}, {Wiersema}, {Bauer}, {Bensch}, {Campana}, {Cano},
  {Covino}, {Coward}, {De Cia}, {de Gregorio-Monsalvo}, {De Pasquale}, {Fynbo},
  {Greiner}, {Gomboc}, {Hanlon}, {Hansen}, {Hartmann}, {Heintz}, {Jakobsson},
  {Kobayashi}, {Malesani}, {Martone}, {Meintjes}, {Micha{\l}owski}, {Mundell},
  {Murphy}, {Oates}, {Salmon}, {van Soelen}, {Tanvir}, {Turpin}, {Xu}, \&
  {Zafar}}]{deUgartePostigo2018}
{de Ugarte Postigo}, A., {Th{\"o}ne}, C.~C., {Bolmer}, J., {et~al.} 2018, \aap,
  620, A119

\bibitem[{{Downes}(1989)}]{Downes1989}
{Downes}, D. 1989, in Lecture Notes in Physics, Berlin Springer Verlag, Vol.
  333, Evolution of Galaxies: Astronomical Observations, ed. {I.~Appenzeller,
  H.~J.~Habing, \& P.~Lena}, 351

\bibitem[{{Endres} {et~al.}(2016){Endres}, {Schlemmer}, {Schilke}, {Stutzki},
  \& {M{\"u}ller}}]{Endres2016}
{Endres}, C.~P., {Schlemmer}, S., {Schilke}, P., {Stutzki}, J., \&
  {M{\"u}ller}, H. S.~P. 2016, Journal of Molecular Spectroscopy, 327, 95

\bibitem[{{Goicoechea} {et~al.}(2015){Goicoechea}, {Chavarr{\'{\i}}a},
  {Cernicharo}, {Neufeld}, {Vavrek}, {Bergin}, {Cuadrado}, {Encrenaz},
  {Etxaluze}, {Melnick}, \& {Polehampton}}]{Goicoechea2015}
{Goicoechea}, J.~R., {Chavarr{\'{\i}}a}, L., {Cernicharo}, J., {et~al.} 2015,
  \apj, 799, 102

\bibitem[{{Goldsmith} \& {Langer}(1999)}]{Goldsmith1999}
{Goldsmith}, P.~F. \& {Langer}, W.~D. 1999, \apj, 517, 209

\bibitem[{{Greisen} {et~al.}(2006){Greisen}, {Calabretta}, {Valdes}, \&
  {Allen}}]{Greisen2006}
{Greisen}, E.~W., {Calabretta}, M.~R., {Valdes}, F.~G., \& {Allen}, S.~L. 2006,
  \aap, 446, 747

\bibitem[{{Harada} {et~al.}(2018){Harada}, {Sakamoto}, {Mart{\'{\i}}n},
  {Aalto}, {Aladro}, \& {Sliwa}}]{Harada2018}
{Harada}, N., {Sakamoto}, K., {Mart{\'{\i}}n}, S., {et~al.} 2018, \apj, 855, 49

\bibitem[{{Henkel} {et~al.}(1998){Henkel}, {Chin}, {Mauersberger}, \&
  {Whiteoak}}]{Henkel1998}
{Henkel}, C., {Chin}, Y., {Mauersberger}, R., \& {Whiteoak}, J.~B. 1998, \aap,
  329, 443

\bibitem[{{Jim{\'e}nez-Serra} {et~al.}(2016){Jim{\'e}nez-Serra}, {Vasyunin},
  {Caselli}, {Marcelino}, {Billot}, {Viti}, {Testi}, {Vastel}, {Lefloch}, \&
  {Bachiller}}]{Jimenez-Serra2016}
{Jim{\'e}nez-Serra}, I., {Vasyunin}, A.~I., {Caselli}, P., {et~al.} 2016,
  \apjl, 830, L6

\bibitem[{{Joye} \& {Mandel}(2003)}]{Joye2003}
{Joye}, W.~A. \& {Mandel}, E. 2003, in Astronomical Society of the Pacific
  Conference Series, Vol. 295, Astronomical Data Analysis Software and Systems
  XII, ed. H.~E. {Payne}, R.~I. {Jedrzejewski}, \& R.~N. {Hook}, 489

\bibitem[{Levenberg(1944)}]{levenberg1944}
Levenberg, K. 1944, Quarterly of Applied Mathematics, 2, 164

\bibitem[{Lovas(1984)}]{Lovas1984}
Lovas. 1984, Spectral Line Atlas of Interstellar Molecules, (Gaithersburg:
  National Inst. Standards and Technology) Magnetic Tape version T84

\bibitem[{{Lovas}(1992)}]{Lovas1992}
{Lovas}, F.~J. 1992, Journal of Physical and Chemical Reference Data, 21, 181

\bibitem[{{Lovas}(2004)}]{Lovas2004}
{Lovas}, F.~J. 2004, Journal of Physical and Chemical Reference Data, 33, 177

\bibitem[{{Mangum} \& {Shirley}(2015)}]{Mangum2015}
{Mangum}, J.~G. \& {Shirley}, Y.~L. 2015, \pasp, 127, 266

\bibitem[{{Maret} {et~al.}(2011){Maret}, {Hily-Blant}, {Pety}, {Bardeau}, \&
  {Reynier}}]{Maret2011}
{Maret}, S., {Hily-Blant}, P., {Pety}, J., {Bardeau}, S., \& {Reynier}, E.
  2011, \aap, 526, A47

\bibitem[{Marquardt(1963)}]{marquardt1963}
Marquardt, D.~W. 1963, SIAM Journal on Applied Mathematics, 11, 431

\bibitem[{{Mart{\'{\i}}n} {et~al.}(2015){Mart{\'{\i}}n}, {Kohno}, {Izumi},
  {Krips}, {Meier}, {Aladro}, {Matsushita}, {Takano}, {Turner}, {Espada},
  {Nakajima}, {Terashima}, {Fathi}, {Hsieh}, {Imanishi}, {Lundgren}, {Nakai},
  {Schinnerer}, {Sheth}, \& {Wiklind}}]{Martin2015}
{Mart{\'{\i}}n}, S., {Kohno}, K., {Izumi}, T., {et~al.} 2015, \aap, 573, A116

\bibitem[{{Mart{\'{\i}}n} {et~al.}(2006){Mart{\'{\i}}n}, {Mauersberger},
  {Mart{\'{\i}}n-Pintado}, {Henkel}, \& {Garc{\'{\i}}a-Burillo}}]{Mart'in2006}
{Mart{\'{\i}}n}, S., {Mauersberger}, R., {Mart{\'{\i}}n-Pintado}, J., {Henkel},
  C., \& {Garc{\'{\i}}a-Burillo}, S. 2006, \apjs, 164, 450

\bibitem[{{Mart{\'\i}n} {et~al.}(2019){Mart{\'\i}n}, {Muller}, {Henkel},
  {Meier}, {Aladro}, {Sakamoto}, \& {van der Werf}}]{Martin2019}
{Mart{\'\i}n}, S., {Muller}, S., {Henkel}, C., {et~al.} 2019, \aap, 624, A125

\bibitem[{{Mart{\'{\i}}n} {et~al.}(2014){Mart{\'{\i}}n}, {Verdes-Montenegro},
  {Aladro}, {Espada}, {Argudo-Fern{\'a}ndez}, {Kramer}, \&
  {Scott}}]{Martin2014}
{Mart{\'{\i}}n}, S., {Verdes-Montenegro}, L., {Aladro}, R., {et~al.} 2014,
  \aap, 563, L6

\bibitem[{{Mart{\'{\i}}n-Dom{\'e}nech}
  {et~al.}(2017){Mart{\'{\i}}n-Dom{\'e}nech}, {Rivilla}, {Jim{\'e}nez-Serra},
  {Qu{\'e}nard}, {Testi}, \& {Mart{\'{\i}}n-Pintado}}]{Martin-Domenech2017}
{Mart{\'{\i}}n-Dom{\'e}nech}, R., {Rivilla}, V.~M., {Jim{\'e}nez-Serra}, I.,
  {et~al.} 2017, \mnras, 469, 2230

\bibitem[{{Martin-Pintado} {et~al.}(1985){Martin-Pintado}, {Wilson}, {Gardner},
  \& {Henkel}}]{Martin-Pintado1985}
{Martin-Pintado}, J., {Wilson}, T.~L., {Gardner}, F.~F., \& {Henkel}, C. 1985,
  \aap, 142, 131

\bibitem[{{McMullin} {et~al.}(2007){McMullin}, {Waters}, {Schiebel}, {Young},
  \& {Golap}}]{McMullin2007}
{McMullin}, J.~P., {Waters}, B., {Schiebel}, D., {Young}, W., \& {Golap}, K.
  2007, in Astronomical Society of the Pacific Conference Series, Vol. 376,
  Astronomical Data Analysis Software and Systems XVI, ed. R.~A. {Shaw},
  F.~{Hill}, \& D.~J. {Bell}, 127

\bibitem[{{M{\"o}ller} {et~al.}(2017){M{\"o}ller}, {Endres}, \&
  {Schilke}}]{Moeller2017}
{M{\"o}ller}, T., {Endres}, C., \& {Schilke}, P. 2017, \aap, 598, A7

\bibitem[{{Moscadelli} {et~al.}(2018){Moscadelli}, {Rivilla}, {Cesaroni},
  {Beltr{\'a}n}, {S{\'a}nchez-Monge}, {Schilke}, {Mottram}, {Ahmadi}, {Allen},
  {Beuther}, {Csengeri}, {Etoka}, {Galli}, {Goddi}, {Johnston}, {Klaassen},
  {Kuiper}, {Kumar}, {Maud}, {M{\"o}ller}, {Peters}, {Van der Tak}, \&
  {Vig}}]{Moscadelli2018}
{Moscadelli}, L., {Rivilla}, V.~M., {Cesaroni}, R., {et~al.} 2018, \aap, 616,
  A66

\bibitem[{{M{\"u}ller} {et~al.}(2005){M{\"u}ller}, {Schl{\"o}der}, {Stutzki},
  \& {Winnewisser}}]{Muller2005}
{M{\"u}ller}, H.~S.~P., {Schl{\"o}der}, F., {Stutzki}, J., \& {Winnewisser}, G.
  2005, Journal of Molecular Structure, 742, 215

\bibitem[{{M{\"u}ller} {et~al.}(2001){M{\"u}ller}, {Thorwirth}, {Roth}, \&
  {Winnewisser}}]{Muller2001}
{M{\"u}ller}, H.~S.~P., {Thorwirth}, S., {Roth}, D.~A., \& {Winnewisser}, G.
  2001, \aap, 370, L49

\bibitem[{{Muller} {et~al.}(2011){Muller}, {Beelen}, {Gu{\'e}lin}, {Aalto},
  {Black}, {Combes}, {Curran}, {Theule}, \& {Longmore}}]{Muller2011}
{Muller}, S., {Beelen}, A., {Gu{\'e}lin}, M., {et~al.} 2011, \aap, 535, A103

\bibitem[{{Pickett} {et~al.}(1998){Pickett}, {Poynter}, {Cohen}, {Delitsky},
  {Pearson}, \& {Muller}}]{Pickett1998}
{Pickett}, H.~M., {Poynter}, I.~R.~L., {Cohen}, E.~A., {et~al.} 1998, Journal
  of Quantitative Spectroscopy and Radiative Transfer, 60, 883

\bibitem[{Press {et~al.}(2007)Press, Teukolsky, Vetterling, \&
  Flannery}]{press2007}
Press, W.~H., Teukolsky, S.~A., Vetterling, W.~T., \& Flannery, B.~P. 2007,
  Numerical Recipes 3rd Edition: The Art of Scientific Computing, 3rd edn. (New
  York, NY, USA: Cambridge University Press)

\bibitem[{{Riquelme} {et~al.}(2018){Riquelme}, {Amo-Baladr{\'o}n},
  {Mart{\'{\i}}n-Pintado}, {Mauersberger}, {Mart{\'{\i}}n}, {Burton},
  {Cunningham}, {Jones}, {Menten}, {Bronfman}, \& {G{\"u}sten}}]{Riquelme2018}
{Riquelme}, D., {Amo-Baladr{\'o}n}, M.~A., {Mart{\'{\i}}n-Pintado}, J.,
  {et~al.} 2018, \aap, 613, A42

\bibitem[{{Rivilla} {et~al.}(2017{\natexlab{a}}){Rivilla}, {Beltr{\'a}n},
  {Cesaroni}, {Fontani}, {Codella}, \& {Zhang}}]{Rivilla2017a}
{Rivilla}, V.~M., {Beltr{\'a}n}, M.~T., {Cesaroni}, R., {et~al.}
  2017{\natexlab{a}}, \aap, 598, A59

\bibitem[{{Rivilla} {et~al.}(2017{\natexlab{b}}){Rivilla}, {Beltr{\'a}n},
  {Mart{\'{\i}}n-Pintado}, {Fontani}, {Caselli}, \& {Cesaroni}}]{Rivilla2017}
{Rivilla}, V.~M., {Beltr{\'a}n}, M.~T., {Mart{\'{\i}}n-Pintado}, J., {et~al.}
  2017{\natexlab{b}}, \aap, 599, A26

\bibitem[{{Rivilla} {et~al.}(2019{\natexlab{a}}){Rivilla}, {Beltr{\'a}n},
  {Vasyunin}, {Caselli}, {Viti}, {Fontani}, \& {Cesaroni}}]{Rivilla2019}
{Rivilla}, V.~M., {Beltr{\'a}n}, M.~T., {Vasyunin}, A., {et~al.}
  2019{\natexlab{a}}, \mnras, 483, 806

\bibitem[{{Rivilla} {et~al.}(2016){Rivilla}, {Fontani}, {Beltr{\'a}n},
  {Vasyunin}, {Caselli}, {Mart{\'{\i}}n-Pintado}, \& {Cesaroni}}]{Rivilla2016}
{Rivilla}, V.~M., {Fontani}, F., {Beltr{\'a}n}, M.~T., {et~al.} 2016, \apj,
  826, 161

\bibitem[{{Rivilla} {et~al.}(2018){Rivilla}, {Jim{\'e}nez-Serra}, {Zeng},
  {Mart{\'{\i}}n}, {Mart{\'{\i}}n-Pintado}, {Armijos-Abenda{\~n}o}, {Viti},
  {Aladro}, {Riquelme}, {Requena-Torres}, {Qu{\'e}nard}, {Fontani}, \&
  {Beltr{\'a}n}}]{Rivilla2018}
{Rivilla}, V.~M., {Jim{\'e}nez-Serra}, I., {Zeng}, S., {et~al.} 2018, \mnras,
  475, L30

\bibitem[{{Rivilla} {et~al.}(2019{\natexlab{b}}){Rivilla},
  {Mart{\'\i}n-Pintado}, {Jim{\'e}nez-Serra}, {Zeng}, {Mart{\'\i}n},
  {Armijos-Abenda{\~n}o}, {Requena-Torres}, {Aladro}, \&
  {Riquelme}}]{Rivilla2019a}
{Rivilla}, V.~M., {Mart{\'\i}n-Pintado}, J., {Jim{\'e}nez-Serra}, I., {et~al.}
  2019{\natexlab{b}}, \mnras, 483, L114

\bibitem[{{Rizzo} {et~al.}(2017){Rizzo}, {Tercero}, \&
  {Cernicharo}}]{Rizzo2017}
{Rizzo}, J.~R., {Tercero}, B., \& {Cernicharo}, J. 2017, \aap, 605, A76

\bibitem[{Schindelin {et~al.}(2015)Schindelin, Rueden, Hiner, \&
  Eliceiri}]{Schindelin2015}
Schindelin, J., Rueden, C.~T., Hiner, M.~C., \& Eliceiri, K.~W. 2015, Molecular
  Reproduction and Development, 82, 518

\bibitem[{Schneider {et~al.}(2012)Schneider, Rasband, \&
  Eliceiri}]{Schneider2012}
Schneider, C.~A., Rasband, W.~S., \& Eliceiri, K.~W. 2012, Nature Methods, 9,
  671

\bibitem[{{Sewi{\l}o} {et~al.}(2018){Sewi{\l}o}, {Indebetouw}, {Charnley},
  {Zahorecz}, {Oliveira}, {van Loon}, {Ward}, {Chen}, {Wiseman}, {Fukui},
  {Kawamura}, {Meixner}, {Onishi}, \& {Schilke}}]{Sewilo2018}
{Sewi{\l}o}, M., {Indebetouw}, R., {Charnley}, S.~B., {et~al.} 2018, \apjl,
  853, L19

\bibitem[{{Towle} {et~al.}(1996){Towle}, {Feldman}, \& {Watson}}]{Towle1996}
{Towle}, J.~P., {Feldman}, P.~A., \& {Watson}, J.~K.~G. 1996, \apjs, 107, 747

\bibitem[{{van der Tak} {et~al.}(2007){van der Tak}, {Black}, {Sch{\"o}ier},
  {Jansen}, \& {van Dishoeck}}]{vandertak2007}
{van der Tak}, F.~F.~S., {Black}, J.~H., {Sch{\"o}ier}, F.~L., {Jansen}, D.~J.,
  \& {van Dishoeck}, E.~F. 2007, \aap, 468, 627

\bibitem[{{Zahorecz} {et~al.}(2017){Zahorecz}, {Jimenez-Serra}, {Testi},
  {Immer}, {Fontani}, {Caselli}, {Wang}, \& {Toth}}]{Zahorecz2017}
{Zahorecz}, S., {Jimenez-Serra}, I., {Testi}, L., {et~al.} 2017, \aap, 602, L3

\bibitem[{{Zeng} {et~al.}(2018){Zeng}, {Jim{\'e}nez-Serra}, {Rivilla},
  {Mart{\'{\i}}n}, {Mart{\'{\i}}n-Pintado}, {Requena-Torres},
  {Armijos-Abenda{\~n}o}, {Riquelme}, \& {Aladro}}]{Zeng2018}
{Zeng}, S., {Jim{\'e}nez-Serra}, I., {Rivilla}, V.~M., {et~al.} 2018, \mnras,
  478, 2962

\end{thebibliography}

\Online

\begin{appendix} 



\section{Units included in MADCUBA}
\label{sec.units}

In this section we present a comprehensive list of the units managed by MADCUBA.

\subsection {Spectral units} 
We refer the reader to \citet{Greisen2006} for a complete description of the spectral coordinates representation and conversion between the units below. 
\begin{itemize}
\item[\textbullet]Frequency (rest and observed): $\textbf{Hz}, \textbf{KHz}, \textbf{MHz}, \textbf{GHz}, \textbf{THz}$
\item[\textbullet]Wavenumbers: $\bf cm^{-1},m^{-1}$
\item[\textbullet]Energy: $\textbf{eV}, \textbf{J}, \textbf{ergs}$
\item[\textbullet]Wavelength (rest and observed): {\bf \AA, nm, {\boldmath$\mu$}m, mm, m}
\item[\textbullet]Velocity (radio, optical, and relativistic): {$\bf m~s^{-1}, km~s^{-1}$}
\item[\textbullet]Redshift: $\bf z $ 
\item[\textbullet]Beta factor: {\boldmath$\beta =v/c$}

\end{itemize}

%
\subsection {Intensity units}
\begin{itemize}
\item[\textbullet][{\boldmath$\mu $} | {\bf m}]  {\bf K}
\item[\textbullet][{\boldmath$\mu $} | {\bf m}]{\bf Jy beam$^{-1}$}
\item[\textbullet][{\boldmath$\mu $} | {\bf m}] {\bf Jy arcsec$^{-2}$}
\item[\textbullet][{\boldmath$\mu $} | {\bf m | M}] {\bf Jy sr$^{-1}$}
\item[\textbullet][{\boldmath$\mu $} | {\bf m}]  {\bf K spaxel$^{-1}$}
\item[\textbullet]{\bf W} [{\bf cm | m}]$\bf^{-2} ~ Hz^{-1} ~ sr^{-1}$ 
\item[\textbullet]{\bf W} [{\bf cm | m}]$\bf^{-2} ~$[{\bf \AA~ | nm | {\boldmath$\mu$}m}]$\bf^{-1}$ {\bf sr$\bf^{-1}$}
\item[\textbullet]$\bf ergs~ s^{-1}~ cm^{-2}~ Hz^{-1}~ sr^{-1} $
\item[\textbullet]$\bf ergs~ s^{-1}~ cm^{-2}$ [{\bf \AA~ | nm | {\boldmath$\mu$}m}]$\bf^{-1}$ $\bf sr^{-1}$
\end{itemize}

\subsection {Flux density units}

\begin{itemize}
\item[\textbullet][{\boldmath$\mu $} | {\bf m | M}] {\bf Jy}
\item[\textbullet]{\bf W } [{\bf cm | m}]$\bf^{-2}$ $\bf Hz^{-1}$ 
\item[\textbullet]{\bf W} [{\bf cm | m}]$\bf^{-2}$ [{\bf \AA~ | nm | {\boldmath$\mu$}m}]$\bf^{-1}$
\item[\textbullet]$\bf ergs~ s^{-1}$ [{\bf cm | m}]$\bf^{-2}$ $\bf Hz^{-1}$ 
\item[\textbullet]$\bf ergs~ s^{-1}$ [{\bf cm | m}]$\bf^{-2}$ [{\bf \AA~ | nm | {\boldmath$\mu$}m}]$\bf^{-1}$
\end{itemize}

\subsection {Integrated flux density units}
\begin{itemize}
\item[\textbullet][{\boldmath$\mu $} | {\bf m | M}] {\bf Jy Hz} 
\item[\textbullet]{\bf W} [{\bf cm | m}]$\bf^{-2}$ 
\item[\textbullet]$\bf ergs~ s^{-1}~ cm^{-2}$
\end{itemize}

\subsection {Integrated Intensity units}
\begin{itemize}
\item[\textbullet]{\bf K Hz} 
\item[\textbullet][{\boldmath$\mu $} | {\bf m}] {\bf K} [{\bf km | m}] $\bf s^{-1}$
\item[\textbullet][{\boldmath$\mu $} | {\bf m}] {\bf Jy beam$\bf^{-1}$} [{\bf km | m}] $\bf s^{-1}$
\item[\textbullet][{\boldmath$\mu $} | {\bf m}] {\bf Jy arcsec$\bf^{-2}$} [{\bf km | m}] $\bf s^{-1}$
\item[\textbullet][{\boldmath$\mu $} | {\bf m | M}] {\bf Jy sr$\bf^{-1}$}  [{\bf km | m}] $\bf s^{-1}$
\item[\textbullet][{\boldmath$\mu $} | {\bf m}] {\bf Jy spaxel$\bf^{-1}$} [{\bf km | m}] $\bf s^{-1}$
\item[\textbullet]{\bf W} [{\bf cm | m}]$\bf^{-2}~ sr^{-1}$
\item[\textbullet]$\bf erg s^{-1}~ cm^{-2}~ sr^{-1}$
\end{itemize}

\section{MADCUBA sample files}
Although we do refer to the documentation which is developing in the MADCUBA webpage\footnote{\url{http://cab.inta-csic.es/madcuba/MADCUBA_IMAGEJ/ImageJMadcuba_Documentation.html}}, in this appendix we include some sample of input files to illustrate the format required by MADCUBA and SLIM.
\subsection{Input Spectra in ASCII format}
\label{sec.sampleasciiformat}
In order to import spectra from any data source, it is possible to resource to a simple ASCII file where the key spectral parameters can be provided as the following sample file. 

\begin{adjustbox}{max width=0.5\textwidth}
\begin{lstlisting}
UnitAngle: deg
UnitSpectral: Hz 
UnitVelo: km/s  
UnitInten: K  
TempScale: TA* 

Beammaj: 0.002777778 
Beammin: 0.001388889 
BeamPA: 20 

CoordType= EQU 
CoordProj:GLS  
Epoch= 2000  
sRadesys=FK5"
XCoord= 66.666   
yCoord= -10.1010 

VeloType: VRAD-LSR
Velocity= 40
xLabel:FREQ-LSR
RestFreq: 8.8E+10
ylabel: Intensity  

// Data x_values1, y_values1, error1
8.7992763427734E+10 -0.014919484034181
8.7994763427734E+10 -0.016418563202024
8.7996763427734E+10 -0.011609123088419
8.7998763427734E+10 -4.1951001621783E-03
8.8000763427734E+10 -1.797300792532E-04
8.8002763427734E+10 -6.2841214239597E-03
\end{lstlisting}
\end{adjustbox}

Note that the importance of the unit definition at the top of the file which will determine the input unit of the different parameters.
However, it is not required to provide values for all parameters. The sample below, does define the basic most header that includes the spectral units and intensity units, and within a single file inputs fours spectra that MADCUBA will load as individual spectra:

\begin{adjustbox}{max width=0.5\textwidth}
\begin{lstlisting}
# mycube_band4_spw1.image-raster
UnitSpectral: GHz 
UnitInten: Jy/beam
9.983684539795e+01 2.884732144998e-03
9.980559539795e+01 2.525573612008e-03
9.977434539795e+01 2.476226447418e-03

# mycube_band4_spw2.image-raster
UnitSpectral: GHz 
UnitInten: Jy/beam
1.074863204956e+02 3.165612609541e-03
1.075175704956e+02 3.662855322292e-03
1.075488204956e+02 3.426149531019e-03

# mycube_band6_spw1.image-raster
UnitSpectral: GHz 
UnitInten: Jy/beam
1.056113967896e+02 3.483511136527e-03
1.056426467896e+02 2.839534894392e-03
1.056738967896e+02 3.722954211636e-03

# mycube_band6_spw2.image-raster
UnitSpectral: GHz 
UnitInten: Jy/beam
9.733055114746e+01 2.297109951407e-03
9.729930114746e+01 3.345324198354e-03
9.726805114746e+01 1.939967128391e-03
\end{lstlisting}
\end{adjustbox}

Note that the lines defining the unit will be used to demarcate individual concatenated spectra within a single input file. For each spectra in the file any non defined parameter will be set to the default units.

\subsection{Sample MADCUBA script}
\label{sec.samplemacro}

The best way to create scripts (macros) in ImageJ and therefore in MADCUBA is through the macro recorder that can be accessed through the Macros menu in MADCUBA or directly through Plugins>Macros>Record within the ImageJ interface.
The file below, created through the macro recorder, shows a sample script that imports three ALMA data cubes, close them, open the imported MADCUBA created fits cubes, extract the spectra from the two regions of interest (in this case, a pixel and a rectangle defined in the reference cube) and writes it to an output spectra file.\\

\begin{adjustbox}{max width=0.5\textwidth}
\begin{lstlisting}
// Import the casa cubes into MADCUBA fits files, changing the velocity
run("Import ALMA CUBE FITS FILE", "select='/path/spw1.cube.fits' changehead='Vrad$180$km/s#'");
selectWindow("PLOT MAD_CUB_spw1.cube.fits");
close();
selectWindow("CUBE MAD_CUB_spw1.cube.fits");
close();
run("Import ALMA CUBE FITS FILE", "select='/path/spw2.cube.fits' changehead='Vrad$180$km/s#'");
selectWindow("PLOT MAD_CUB_spw1.cube.fits");
close();
selectWindow("CUBE MAD_CUB_spw1.cube.fits");
close();
run("Import ALMA CUBE FITS FILE", "select='/path/spw3.cube.fits' changehead='Vrad$180$km/s#'");
selectWindow("PLOT MAD_CUB_spw1.cube.fits");
close();
selectWindow("CUBE MAD_CUB_spw1.cube.fits");
close();
// Open the Master cube, which will be the reference for all the others.
run("Open Virtual Cube", "writemacro=true select='/mypath/MAD_CUB_spw1.cube.fits'");
// Open all other cubes
run("Open Virtual Cube Background", "select='/mypath/MAD_CUB_spw2.cube.fits'");
run("Open Virtual Cube Background", "select='/mypath/MAD_CUB_spw3.cube.fits'");
run("Synchronize Cube", "");
//INFO: Synchronize List Selected Cube
run("Synchronize Select Cube", "name_cube=ALL");
selectWindow("CUBE MAD_CUB_spw1.cube.fits");
// Select the Region of Interest (ROI)
// Extracting spectra in a single pixel
makeRectangle(69, 96, 1, 1);
run("Synchronize Roi Cube", "cursorroi=true coords=true");
run("Synchronize Cube", "");
run("Extract Spectra Sync Integrated", "select='/mypath/singlepixelROI.fits'");
// EXTRACTING POSITION2
makeRectangle(69, 96, 10, 10);
run("Synchronize Roi Cube", "cursorroi=true coords=true");
run("Synchronize Cube", "");
run("Extract Spectra Sync Integrated", "select='/mypath/rectangleROI.fits'");
close();
\end{lstlisting}
\end{adjustbox}

\subsection{Continuum by user input file}
\label{sec.samplecontinuuminput}
The continuum of a spectrum can be added as an user input in a simple two column file consisting of the frequency in Hz and the intensity, in the units of the current open spectra,  at that frequency.
Here is a simple sample continuum input file.

\begin{adjustbox}{max width=0.5\textwidth}
\begin{lstlisting}
604.0E+09,0.0018
671.0E+09,0.0027
758.0E+09,0.0040
903.0E+09,0.0066
995.0E+09,0.0085
1025.0E+09,0.0096
1043.0E+09,0.0104
\end{lstlisting}
\end{adjustbox}

Note that continuum values will be interpolated in between the values in the file.

\onecolumn
\section{Detailed derivative equations for fitting algorithms}
\label{sec.derivatives}

Here we present the analytic derivative equations of the Eq.~\ref{eq.unitssizes} used to generate the Design matrix (Sec.~\ref{sec.fitting}).

For the sake of simplicity we will rewrite Eq.~\ref{eq.summatorylinetemperature} to represent the intensity measured in the corresponding units for a single molecule and component as

\begin{equation}
I = U_f~J_{TOT}~ (1-e^{-\sum_t \tau_\circ^t~\Phi})
\end{equation}
where we have defined the Gaussian shape ($\Phi$) and the total intensity in temperature units ($J_{TOT}$) as
\begin{align}
\Phi       & = e^{-4\,ln 2({\rm v}-{\rm v}_\circ)^2/\Delta {\rm v}^2} \\
J_{TOT}    & = \bigl( J_\nu(T_{ex})-J_\nu(T_{bg})-J_\nu(T_{c}) \bigr)
\end{align}
and the unit factor $U_f$ will be either $\frac{\theta_{s}^2 }{\theta_{s}^2 +\theta_{b}^2}$  (for units of main beam temperature $T_{MB}$) or $10^{23} ~\frac{2k\nu^2}{c^2} 1.133~\theta_{s}^2$ (for units of flux density $S$), as introduced in Eq.~\ref{eq.unitssizes}.

\begin{align}
\frac{\delta I}{\delta N}  =~ &  U_f~J_{TOT}~\bigl( 
                             e^{-\sum_i~\tau_\circ^i~\Phi } 
                             \frac{1}{N}~\sum_i~\tau_\circ^i~\Phi
                            \bigr) \\
\frac{\delta I}{\delta T_{ex}} =~ &  
                           U_f~\biggl( J(T_{ex})^2~\frac{e^{h\nu/kT_{ex}}}{T_{ex}^2}~(1-e^{-\sum_i~\tau_\circ~\Phi } ) + \\
                           & + J_{TOT}~\Bigl(
                             e^{-\sum_i~x\tau_\circ^i~\Phi } \sum_i~\tau_\circ^i~\Phi
                             \bigl(\frac{(E_l~e^{-E_l/kT_{ex}}-E_u~e^{-E_u/kT_{ex}})}{k~T_{ex}^2~(e^{-E_l/kT_{ex}}-e^{-E_u/kT_{ex}})} - \frac{1}{Q(T_{ex})}~\frac{\delta Q(T_{ex})}{\delta T_{ex}}
                             \bigr)
                            \Bigr)\biggr)\\
\frac{\delta I}{\delta {\rm v}_\circ} =~ & U_f~J_{TOT}~\Bigl(
                             e^{-\sum_i~\tau_\circ^i~\Phi } \sum_i~(
                             \frac{8\,ln 2~({\rm v}-{\rm v}_\circ)}{\Delta {\rm v}^2}~\tau_\circ^i~\Phi
                             )
                            \Bigr)\\
\frac{\delta I}{\delta \Delta {\rm v}} =~ & U_f~ J_{TOT}~\Bigl(
                             e^{-\sum_i~\tau_\circ^i~\Phi } \sum_i~
			     \frac{\tau_\circ^i~\Phi}{\Delta {\rm v}}
                             (\frac{8\,ln 2~({\rm v}-{\rm v}_\circ)^2}{\Delta {\rm v}^2}
                             - 1
                             )
                            \Bigr) \\
\frac{\delta I}{\delta \Delta \theta_s} =~ & U'_f~ J_{TOT}~ (1-e^{-\sum_t \tau_\circ^t~\Phi})                            
\end{align}

where $U'_f=\frac{\delta U_f}{\theta_s}$ will be $\frac{2}{\theta_s}~\Bigl( 
\frac{\theta_s^2}{(\theta_s^2+\theta_b^2)}-\bigl(\frac{\theta_s^2}{(\theta_s^2+\theta_b^2)}\bigr)^2\Bigr)$ in $T_{MB}$ units, and $10^{23}~2.266~\theta_s~\frac{2k\nu^2}{c^2}$ in $S$ units.

In order to derive the partial derivatives of the measured intensity above, the partial derivatives of the central optical depth, Gaussian shape and total intensity are required. For completeness, we write the expression of these derivative below. We note that the partial derivative of the partition function with respect to the excitation temperature ($\frac{\delta Q(T_{ex})}{\delta T_{ex}}$) appearing in $\frac{\delta I}{\delta T_{ex}}$ is approximated in SLIM by the linear slope of the the partition function between the two values from the catalog entries surrounding a given excition temperature.

\begin{align}
\frac{\delta U_f~J_{TOT}}{\delta N} =~ & 0 \\
\frac{\delta U_f~J_{TOT}}{\delta T_{ex}} =~ & U_f~J(T_{ex})^2~\frac{e^{h\nu/kT_{ex}}}{T_{ex}^2} \\
\frac{\delta U_f~J_{TOT}}{\delta {\rm v}_\circ} =~ & 0 \\
\frac{\delta U_f~J_{TOT}}{\delta \Delta {\rm v}} =~ & 0 \\
\frac{\delta U_f~J_{TOT}}{\delta \theta_s} =~ & U'_f~J_{TOT} \\
\end{align}

\begin{align}
\frac{\delta \tau_\circ}{\delta N} =~ & \frac{\tau_\circ}{N} \\
\frac{\delta \tau_\circ}{\delta T_{ex}} =~ &   \tau_\circ ~\Bigl(  
               \frac{(E_l~e^{-E_l/kT_{ex}}-E_u~e^{-E_u/kT_{ex}})}{k~T_{ex}^2~(e^{-E_l/kT_{ex}}-e^{-E_u/kT_{ex}})} - \frac{1}{Q(T_{ex})}~\frac{\delta Q(T_{ex}}{\delta T_{ex}})  \Bigr) \\
\frac{\delta \tau_\circ}{\delta {\rm v}_\circ} =~ & 0 \\
\frac{\delta \tau_\circ}{\delta \Delta {\rm v}} =~ &  - \frac{\tau_\circ}{\Delta {\rm v}} \\
\frac{\delta \tau_\circ}{\delta \theta_s} =~ & 0 \\
\end{align}

\begin{align}
\frac{\delta \Phi}{\delta N} =~ & 0 \\
\frac{\delta \Phi}{\delta T_{ex}} =~ & 0 \\
\frac{\delta \Phi}{\delta {\rm v}_\circ} =~ & \frac{8\,ln 2~({\rm v}-{\rm v}_\circ)}{\Delta {\rm v}^2}~\Phi \\
\frac{\delta \Phi}{\delta \Delta {\rm v}} =~ &\frac{8\,ln 2~({\rm v}-{\rm v}_\circ)^2}{\Delta {\rm v}^3}~\Phi \\
\frac{\delta \Phi}{\delta \theta_s} =~ & 0 \\
\end{align}

\end{appendix}

\end{document}